\RequirePackage{fix-cm}
\documentclass[smallextended]{svjour3}       
\smartqed  
\usepackage{graphicx}
\usepackage{multirow}
\usepackage{algorithm}
\usepackage{algorithmic}
\usepackage{amssymb}
\usepackage{array}
\usepackage{amsmath}
\usepackage{color}
\usepackage[misc]{ifsym} 
\begin{document}

\title{Hierarchical Heuristic Learning towards Efficient Norm Emergence
}


\author{Tianpei Yang         \and
        Jianye Hao     \and
        Zhaopeng Meng \and
        Sandip Sen \and
        Sheng Jin
}


\institute{Tianpei Yang \and Jianye Hao (\Letter)  \and Sheng Jin \at
              School of Computer Software, Tianjin University, China \\              
              \email{jianye.hao@tju.edu.cn}                 
              \and           
              Tianpei Yang  \at
              \email{tpyang@tju.edu.cn}           
              \and      
           Zhaopeng Meng \at
            Tianjin University of Traditional Chinese Medicine, China\\
              \email{mengzp@tju.edu.cn}                              
           \and                    
           Sandip Sen \at
              University of Tulsa, USA\\
              \email{sandip-sen@utulsa.edu}                
              \and 
              Sheng Jin \\
               \email{jinsheng@tju.edu.cn}
}

\date{Received: date / Accepted: date}

\maketitle

\begin{abstract}
Social norms serve as an important mechanism to regulate the behaviors of agents and to facilitate coordination among them in multiagent systems. One important research question is how a norm can rapidly emerge through repeated local interaction within an agent society under different environments when their coordination space becomes large. To address this problem, we propose a Hierarchically Heuristic Learning Strategy (HHLS) under the hierarchical social learning framework, in which subordinate agents report their information to their supervisors, while supervisors can generate instructions (rules and suggestions) based on the information collected from their subordinates. Subordinate agents heuristically update their strategies based on both their own experience and the instructions from their supervisors. Extensive experiment evaluations show that HHLS can support the emergence of desirable social norms more efficiently and is applicable in a much wider range of multiagent interaction scenarios compared with previous work. We also investigate the effectiveness of HHLS by separating out the different components of the HHLS and evaluating the relative importance of those components. The influence of key related factors (e.g., hierarchical factors, non-hierarchical factors, fixed-strategy agents) are investigated as well. 

\end{abstract}

\section{Introduction}
\label{section1}
Social norms play an important role in regulating agents' behaviors to ensure coordination among agents and functioning of agent societies. One commonly adopted characterization of a norm is to model it as a consistent equilibrium that all agents follow during interactions where multiple equivalent equilibria coexist \cite{young1996economics}. How social norms can emerge efficiently in agent societies is a key research problem in the area of normative multiagent systems. 

There exist two major approaches for addressing norm emergence problem: the top-down \cite{Gotnes2010Optimal,Terroine2013Automated} approach and the bottom-up approach \cite{sen2007emergence,yu2013emergence}. The former approach investigates how to efficiently synthesize a norm for all agents beforehand, while the latter one investigates how a norm can emerge through repeated local interactions. In distributed multiagent interaction environments, it is usually difficult to design a norm a priori, and before agents interact, since a centralized controller might not exist and due to the fact that the optimal norm may vary frequently as the environment dynamically changes and therefore, the bottom-up approach of emergent norms via local interactions promises to be more suitable for such kinds of distributed and dynamic environments.

Until now, significant efforts have been devoted to investigating norm emergence problem from the bottom-up approaches \cite{airiau2014emergence,hao2013dynamics,yu2013emergence,bianchi2007heuristic,mihaylov2014decentralized,jianye2015heuristic,zhang2009integrating,kapetanakis2005reinforcement,mukherjee2008norm,sen2010effects,sen2007emergence,shoham1997emergence,villatoro2011social,villatoro2013robust,villatoro2009topology,yu2015hierarchical,yu2016adaptive,yu2016modelling}. Sen and Airiau \cite{sen2007emergence} investigated the norm emergence problem in a population of agents within randomly connected networks where each agent is equipped with certain multiagent learning algorithms. The local interaction among each pair of agents is modeled as two-player normal-form games, and a norm corresponds to one consistent Nash equilibrium of the coordination/anti-coordination game. Later, a number of papers \cite{mukherjee2008norm,villatoro2009topology,DBLP:conf/atal/SenS09,airiau2014emergence} subsequently extended this work by using more realistic and complex networks (e.g., small-world network and scale-free network) to model the diverse interaction patterns among agents. Additionally, different learning strategies and mechanisms have been proposed to better facilitate norm emergence among agents within different interaction environments \cite{zhang2009integrating,savarimuthu2011aspects,mihaylov2014decentralized,bianchi2007heuristic,jianye2015heuristic,yu2016adaptive}. 

Most of the previous works only focus on games of relatively small size. This simplification does not accurately reflect the practical interaction scenarios where the action space of agents can be quite large. With the increase of action space, most of the existing approaches usually result in very slow norm emergence or even fail to converge. Recently, Yu et al. \cite{yu2015hierarchical} proposed a hierarchical learning strategy to improve the norm emergence rate for very large action space problems. However, this work only considers the case in which a norm corresponds to a Nash equilibrium where the interacting agents select the same action. This usually can be modelled as a two-player $n$-action \emph{coordination game} (CG). One simple example with n=2 is shown in Table \ref{table1}. However, in realistic interaction scenarios, a norm may correspond to a Nash equilibrium where all agents select different actions. One notable example is considering two drivers arriving at a road intersection from two neighboring roads. To avoid collision, one possible norm is ``yield to the left", i.e., waiting for the car on the left-hand side to go through the intersection first. This kind of scenario can naturally be modeled as an \emph{anti-coordination game} (ACG), as shown in Table \ref{table2}, which exist two different norms, i.e., (a, b) and (b, a). 

Furthermore, agents may be faced with the challenge of high mis-coordination cost and stochasticity of the environment. One representative example is shown in Table \ref{table3}, which we call \emph{fully stochastic coordination game with high penalty} (FSCGHP). In this game, there exist two optimal Nash equilibria marked with red each of which corresponds to a norm, and one suboptimal Nash equilibrium marked with blue. Two major challenges coexist in this game: agents are vulnerable to converge to the suboptimal Nash equilibrium due to the high penalty when agents mis-coordinate on the outcomes; agents need to effectively distinguish between the stochasticity of the environment and the exploration of other learners. It is not clear, a priori, how a population of agents can efficiently evolve towards a consistent norm, given the large space of possible norms in such challenging environments.

\begin{table}
\centering
\caption{An example of coordination game.} \label{table1}
		\begin{tabular}{cccc}
		\hline 
\rule{0pt}{12pt}
\multirow{2}{0.5 in}{} &  & \multicolumn{2}{c}{Agent 2's actions} \\ 
\cline{3-4}
\rule{0pt}{12pt}
 &&a&b \\ 
\hline
\multirow{2}{0.5 in}{\\Agent 1's \\ actions}\\
 &a&{\color{red}1}&-1 \\ 
\cline{2-4}
\rule{0pt}{12pt}
 &b&-1&{\color{red}1} \\ 
\hline 
		\end{tabular}
\end{table}

\begin{table}
\centering
	\caption{An example of anti-coordination game.} \label{table2}
		\begin{tabular}{cccc}
\hline 
\rule{0pt}{12pt}
\multirow{2}{0.5 in}{}&
& \multicolumn{2}{c}{Agent 2's actions} \\ 
\cline{3-4}
\rule{0pt}{12pt}
 &&a&b \\ 
\hline
\multirow{2}{0.5 in}{\\Agent 1's \\ actions}\\
 &a&-1&{\color{red}1} \\ 
\cline{2-4}
\rule{0pt}{12pt}
 &b&{\color{red}1}&-1 \\ 
\hline 
		\end{tabular}
	\end{table}

\begin{table}
\centering 
\caption{Fully stochastic coordination game with high penalty.} \label{table3}
\begin{tabular}{ccccc}
\hline 
\rule{0pt}{12pt}
\multirow{2}{0.5 in}{1's payoff \\ 2's payoff} &  & \multicolumn{3}{c}{Agent 2's actions} \\ 
\cline{3-5}
\rule{0pt}{12pt}
 & & a & b & c  \\ 
\hline
\multirow{3}{0.5 in}{\\Agent 1's \\ actions}\\
 & a & {\color{red}8/12} & -5/5 & -20/-40 \\ 
\cline{2-5}
\rule{0pt}{12pt}
 & b & -5/5 & {\color{blue}0/14} & -5/5 \\ 
\cline{2-5}
\rule{0pt}{12pt}
 & c & -20/-40 & -5/5 & {\color{red}8/12} \\ 
\hline 
	\end{tabular}
\end{table}
To answer this question, in this paper we propose a novel \underline{H}ierarchical \underline{H}euristic \underline{L}earning \underline{S}trategy (HHLS) under the hierarchical social learning framework to facilitate rapid norm emergence in agent societies. In the hierarchical social learning framework, the agent society is separated into a number of clusters, each of which consists of several subordinate agents. Each cluster's strategies are monitored and guided by one supervisor agent. In each round, each supervisor agent collects the interaction information of its subordinate agents and generates instructions in the forms of behavioral rules and suggestions for its subordinates. On the other hand, each subordinate agent, apart from learning from its local interaction, also adjusts its strategy based on the instructions from its supervisor. The main feature of the proposed framework is that through hierarchical supervision among agents, an effective compromise solution can be generated to effectively balance distributed interactions and centralized control towards efficient and robust norm emergence. 

We evaluate the performance of HHLS under a wide range of games and experimental results show that HHLS can facilitate rapid emergence of norms compared with the state-of-the-art approaches. We investigate the influence of hierarchical (e.g., the cluster size, different grouping mechanisms, the number of disabling supervisors) and non-hierarchical factors (e.g., the population size, the size of action space, different network topologies, the neighborhood size) on norm emergence. Lastly, we investigate the influence of fixed-strategy agents on norm emergence (e.g., initial and late intervention, different placement strategies) and norm emergence in isolated sub-networks. 

The remainder of the paper is organized as follows. Section 2 discusses related work. Section 3 introduces the hierarchical social learning framework and the heuristic learning strategy. Section 4 presents experimental evaluation results compared with two representative state-of-the-art approaches. Finally, Section 5 concludes the paper and points out future directions.

\section{Related Work} \label{section2}
\subsection{Norm Emergence} \label{section2.1}
Norm emergence problem has received a wide range of attention in Multiagent Systems (MASs) literature. Shoham and Tennenholtz \cite{shoham1997emergence} first investigated the norm emergence problem in an agent society based on a simple and natural strategy - the Highest Cumulative Reward (HCR). In this study, they showed that HCR achieved high efficiency on social conventions in a class of games. Sen and Airiau \cite{sen2007emergence} investigated the norm emergence problem in MASs where each agent is equipped with certain existing multiagent learning algorithms. They firstly proposed the model of \emph{social learning}, where each agent learns from repeated interactions with multiple agents in a given scenario. In this study, the local interaction among each pair of agents is modeled as two-player normal-form games, and a norm corresponds to one consistent Nash equilibrium of the game. Later, a number of papers \cite{mukherjee2008norm,villatoro2009topology,sen2010effects,airiau2014emergence} extended this work by leveraging more realistic and complex networks (e.g., small-world network and scale-free network) to model the interaction patterns among agents and evaluated the influence of heterogeneous agent systems and space-constrained interactions on norm emergence. Savarimuthu \cite{savarimuthu2011aspects} recapped the existing mechanisms on the multiagent-based emergence, and investigated the role of three proactive learning methods in accelerating norm emergence. The influence of liars on norm emergence is also considered and simulation results showed that norm emergence can still be sustained in the presence of liars. 

Villatoro et al. \cite{villatoro2009topology} proposed a reward learning mechanism based on interaction history. In this study, they investigated the influence of different network topologies and the effects of memory of past activities on convention emergence. Later, they \cite{villatoro2011social,villatoro2013robust} introduced two rules (i.e., re-wiring links with neighbors and observation) to overcome the suboptimal norm problems. They investigated the influence of Self-Reinforcing Substructure (SRS) in the network on impeding full convergence towards society-wide norms, which usually results in reduced convergence rates. Hao et al. \cite{hao2013dynamics} investigated the problem of coordinating towards optimal joint actions in cooperative games under the social learning framework by introducing two types of learners (IALs and JALs). Yu et al. \cite{yu2013emergence} proposed a novel collective learning framework to investigate the influence of agent local collective behaviors on norm emergence in different scenarios and defined two strategies (collective learning-l and collective learning-g) to promote the emergence of norms where agents are allowed to make collective decisions within networked societies. Later, Hao et al. \cite{jianye2015heuristic} proposed two learning strategies under the collective learning framework: collective learning EV-l and collective learning EV-g to address the problem of high mis-coordination cost and stochasticity in complex and dynamic interaction scenarios. More recently, Yu et al. \cite{yu2016adaptive} proposed an adaptive learning framework for efficient norm emergence. Later, Yu et al. \cite{yu2016modelling} proposed a novel adaptive learning to facilitate consensus formation among agents, in order to efficiently establish a consistent social norm in agent societies. However, all the aforementioned works usually focus on relatively small-size games, and do not address the issue of efficient norm emergence in large action space problems.

Hierarchical learning framework, as a promising solution to accelerate coordination among agents, has been studied in different multiagent applications (e.g., package routing \cite{zhang2009integrating}, traffic control \cite{abdoos2013holonic}, p2p network \cite{campos2011organisational} and smart-grid \cite{ye2011hybrid}). For example, Zhang et al. \cite{zhang2009integrating,zhang2010self} studied the package routing problem and proposed a multi-level organizational structure for automated supervision and a communication protocol for information exchange between higher-level supervising agents and their subordinate agents. Simulation shows that the organization-based control framework can significantly increase the overall package routing efficiency compared to traditional non-hierarchical approaches. Abdoos et al. \cite{abdoos2013holonic} proposed a multi-layer organizational controlling framework to model large traffic networks to improve the coordination between different car agents and improve the overall traffic efficiency. Recently, Yu et al. \cite{yu2015hierarchical} proposed a hierarchical learning framework to study the norm emergence problem. In this study, they proposed a two-level hierarchical framework. Agents in the lower level interact with each other and report information to their supervisors in the higher level, while agents in the higher level called supervisors pass down guidance to the lower level. Agents in the lower level follow guidance in policy update. However, their framework is designed for coordination game only, where agents only need to coordinate to select the same action for effective norm emergence.


\subsection{Fixed-strategy Agents on Norm Emergence} \label{section2.2}
Given the existence of multiple norms, agents usually do not have preference over different norms. It has been found that fixed-strategy agents could play a critical role in influencing the direction of norm emergence and have received wide range of attention in previous work \cite{sen2007emergence,marchant2015manipulating,franks2013manipulating,griffiths2012impact}. Fixed-strategy agents are those who always select the same action regardless of its efficiency or others' choices. Previous work has shown that inserting relatively small numbers of fixed-strategy agents can significantly influence much larger populations when placed in networked social learning framework. 

Sen and Airiau \cite{sen2007emergence,airiau2014emergence} firstly investigated the effect of fixed-strategy agents in affecting norm adoption in social learning framework. Griffiths and Anand \cite{griffiths2012impact} proposed a social learning model where each agent has a fixed length of memory recording the most recent actions it has selected. They demonstrated how fixed-strategy agents can manipulate emergence and evaluated strategies for inserting fixed-strategy agents using placement heuristics such as degree and betweenness centrality. Marchant et al. \cite{marchant2015manipulating} investigated the influence of fixed-strategy agents on norm emergence in a dynamic network. They defined a new heuristic and proposed late intervention of fixed-strategy agents, where the whole system has already emerged an existing norm. Experimental results show that placing small numbers of fixed-strategy agents can disrupt the already established norm and the whole population eventually converged to the norm adopted by the fixed-strategy agents. Tiwari et al. \cite{TiwariJBBG1} investigated the effect of leader placement in simulated robotic swarms. Their experimental results showed that the leader placement strategy determines the time it takes for the swarm to converge: leaders placed in the middle or periphery of the swarm are better in maneuvering the swarm than leaders placed in the front.  

Franks et al. \cite{franks2013manipulating} first introduced \emph{Influencer Agents} (IAs) as a mechanism to manipulate norm emergence direction. The IAs can be considered as agents with uniform levels strategies and goals to aid the emergence of high quality norms in domains characterized by heterogeneous ownership and uniform levels of agent authority. IAs are those agents with specific norms (fixed strategies), such that the whole population, through their rational selection of actions, is guided towards the adoption of the specific norms. They evaluated the influence of IAs in the language coordination problem domain. Results showed that small number of IAs can effectively manipulate the emergence of high-quality conventions. On the other hand, they proved the fragility of convention emergence in the presence of malicious or faulty agents that attempted to propagate low quality conventions, and confirmed the importance of network topologies in norm adoption. Franks et al. \cite{franks2014learning} later proposed a methodology for learning the influence of the placement of agents in a network and evaluated their approach in the context of the language coordination domain. They investigated the influence of IAs in a coordination game with heterogeneous agent learning mechanisms. They used several topological metrics to measure the influence of fixed-strategy agents on norm emergence. Experimental results showed that when placing agents following their methodology, agents can gain much higher influence power than random placement.

Another notable work related with fixed-strategy agents is by Genter et al. \cite{genter2015determining} which investigates the influence of placing fixed-strategy agents in a flock of agents. The group flocking behavior emerges from simple local control rules, through which each individual agent adjusts its own trajectory based on those of its neighbors. They defined several methodologies for placing the influencing agents into the flock. Experimental results showed that the graph approach outperforms other approaches when initially placing influencing agents anywhere into the flock and the grid approach performs best when the agents must travel to their desired positions after being initially placed outside the flock. Later, Genter and Stone \cite{Genter2016Adding} extended their work on considering the placement heuristic of influencing agents in a flock and how influencing agents should behave in order to join a flock in motion (i.e., the joining case). They showed that the placement heuristics of fixed-strategy agents which work well for initial placement also work well if fixed-strategy agents can place themselves in an incoming flock within the flight path. Recently, they \cite{Genter2017Agent} considered the situation where robot birds are deployed to influence flocks in nature. They proposed a hover approach for robot birds to use when joining and leaving a flock. Finally, they summarized and pointed out the main drawback of their hover approach: the robot birds are required to hover at desired positions. This may cause hovering problems because robot birds would be unable to hover when they are recognized by natural birds as `one of their own'. 

\section{Hierarchical Social Learning Framework} \label{section3}
\subsection{Framework Overview} \label{section3.1}
We consider a population of $N$ agents where each agent is connected by the underlying network topology. In each round, each agent interacts with one randomly selected agent from its neighborhood. An agent's neighborhood consists of all agents that are physically connected. We model the interaction between each pair of agents as a normal-form game. At the beginning of each interaction, one agent is randomly assigned as the row player and the other as the column player. We assume that each agent is unaware of opponent's action and payoff information and can only have access to its own action and payoff information during interaction. The population of agents are divided into multiple levels, and the agents in each level supervise the behaviors of agents in its neighboring lower level. 

\begin{figure}
\centerline{\includegraphics[scale = 0.4]{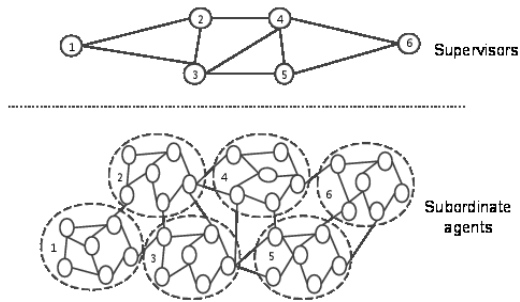}}
\caption{An example of the two-level hierarchical network.} \label{figure1}
\end{figure}

One example of two-level hierarchical network is shown in Fig \ref{figure1}. Each supervisor agent $i$ in the higher level is in charge of a group of subordinate agents (denoted as $sub(i)$) in the bottom level surrounded by dashes lines. For each subordinate agent $j$, its supervisor agent is denoted as $sup(j)$. For subordinates, the topological connections between them are determined by the original network topology; for supervisors, a pair of supervisors are neighboring agents if the corresponding group of subordinates they supervise are connected. Note that each supervisor agent is also allowed to communicate with its neighboring supervisor agents. The way of dividing subordinates into groups and choosing each group's supervisor will be discussed in details in Section \ref{section3.4}.

The interaction protocol of agents under the hierarchical social learning framework is summarized in Algorithm \ref{algorithm1}. In each round, each agent is paired with another agent randomly selected from its neighborhood (Line 3), and their roles are randomly assigned (Line 4). Each agent then chooses an action following its learning strategy (Line 5), and then updates its strategy based on its current-round feedback (Line 6). After that, each subordinate agent reports its action and reward information to its supervisor (Line 7-9). At the end of each round, each supervisor collects all subordinate agents' information, generates and issues the instructions to its subordinate agents (Line 12-14). Finally, each subordinate agent updates its strategy based on the instructions accordingly (Line 15-17).

\begin{algorithm}[h]
\caption{The interaction protocol of the hierarchical framework.} \label{algorithm1}
\begin{algorithmic}[1]
\FOR{each round of interaction}
\FOR{each agent $i\in N$}
\STATE Randomly choose a neighboring agent $j$ to interact; \
\STATE Assign distinct roles randomly $i \rightarrow$ state $s_i$, $j \rightarrow$ state $s_j$
\STATE Select actions $a_{i}$ and $a_j$ and get rewards $r_{i}$ and $r_j$;\
\STATE Update its strategy based on $\langle s_i, a_{i},r_{i}\rangle$.
\IF{agent $i$ is a subordinate agent}
    \STATE Reporting its experience $\langle s_i, a_{i},r_{i}\rangle$ to sup(i);\
\ENDIF
\ENDFOR

\FOR{each supervisor agent j}
\STATE Generate instructions based on the information from $sub(j)$;
\STATE Provide the instructions to $sub(j)$;
\ENDFOR
\FOR{each subordinate agent k}
\STATE Update its strategy based on the instructions from $sup(k)$;\
\ENDFOR
\ENDFOR
\end{algorithmic}
\end{algorithm}

\subsection{Grouping Mechanism} \label{section3.4}
As described in Section \ref{section3.1}, the whole population is divided into two levels. Each agent in the higher level supervises a group of subordinate agents in the lower level. Each supervisor has a dual role in the network. One role is as a supervisor of a group of subordinates and the other role is a subordinate agent which is also allowed to interact with one of its neighbors. How to group subordinate agents and how to choose supervisors are two key aspects that can influence the performance of our algorithm. One natural way is to allow each supervisor to supervise its direct neighbors, which, however, may not be effective due to the sparsity of network topologies. To this end, we propose a number of grouping mechanisms as follows.

\begin{itemize}
\item \emph{Random Grouping} \label{section3.4.1}
\end{itemize}

We divide subordinate agents equally into $k$ groups randomly. The supervisors are also selected randomly from each group. This grouping mechanism is used in our previous work \cite{DBLP:conf/ecai/YangMHSY16}. 

\begin{itemize}
\item \emph{Degree Grouping} \label{section3.4.2}
\end{itemize}

Given any network topology $G \langle V,E \rangle$, each agent $i$ corresponds to a node $v_i \in V$, $N\left(v_i\right)$ denotes the set of neighbors of the agent $i$, and the degree of agent $i$ is $\vert N \left(v_i \right) \vert$. We divided subordinate agents into different groups based on their degrees. Agents with the same degree are allocated into the same group.  Also, the supervisors are randomly selected from each group. Note that degree grouping is not an equally division.

\begin{itemize}
\item \emph{K-means Grouping} \label{section3.4.3}
\end{itemize}

We introduce the core idea of K-means clustering algorithm into our grouping mechanism \cite{macqueen1967some}. In a network $G \langle V,E \rangle$, we define the distance $Dis_{spl}(v_i,v_j)$ between two agents $i$, $j$ in such network as the shortest path between them (i.e. the minimum number of edges connecting these two agents). We first randomly select $k$ agents as the $k$ centering points. Each of the remaining agents is assigned to a group if the distance between it and this center point is minimum over the $k$ centering points (random assignment in case of a tie). The centering agent of each group is recomputed after all agents have been assigned to one group. This process repeats until each agent no longer changes its group. The final centering agent of each group is selected as the supervisor of the group. Note that K-means grouping is also not an equal division and it may lead to extreme cases where some groups have too small number of agents, which causes undesirable results. 

To address this issue, we modify the original K-means grouping to suit our problem domain in which all groups make some adjustments after the original K-means algorithm terminates. We adopt a threshold as the maximum distance between the centering agent and other agents in each group. Agents will be reassigned to other groups if the distance between it and the centering point is larger than the given threshold. Moreover, to prevent a group from being too small, a group will no longer compete for agents if the number of agents in the group size exceeds a maximum threshold.

\begin{itemize}
\item \emph{Kernighan-Lin Grouping} \label{section3.4.4}
\end{itemize}

The Kernighan{-}Lin algorithm is a heuristic algorithm for the graph partitioning problem \cite{kernighan1970efficient}. Subordinate agents are first divided into two equal groups randomly. The sum of the costs of edges between node $v_i \in V$ and other nodes within the same group is denoted as the internal cost of node $v_i \in V$, and the sum of the costs of edges between node $v_i \in V$ and nodes in opposite group is denoted as the external cost of node $v_i \in V$. Each agent makes an exchange from one group to the other if the exchange reduces the internal cost and increases the external cost. Kernighan{-}Lin Grouping ends if no further exchanges are feasible. Each group will be divided into two equal groups to repeat the process described above if we need two more groups. An agent is selected as the supervisor if its degree is the highest in the group.

\subsection{Information Exchange between Supervisors and Subordinates} \label{section3.2}
In the hierarchical social learning framework, subordinate agents send their feedback information to their corresponding supervisors, while supervisors pass down instructions to their corresponding subordinate agents. In details, each subordinate agent $i$ reports its current-round interaction experience $\langle s_i, a_{i},r_{i}\rangle$ to its supervisor $sup(i)$. For supervisors, we distinguish two different forms of instructions that they can provide to their subordinates: $suggestion$ and $rule$ \cite{zhang2009integrating}. Intuitively, a $rule$ is a hard constraint that specifies an action that subordinate agents are forbidden to select under certain state in the next round; in contrast, a $suggestion$ is a soft constraint which indirectly affects the strategies of the subordinate agents next round. 

A set $F$ of $rules$ consists of all the forbidden actions for subordinates under different states. Formally we have, 
\begin{equation} \label{eq1:rule}
F = \lbrace \langle s,a\rangle \vert a \in A, s\in S \rbrace
\end{equation}
where each element $\langle s,a\rangle$ denotes that action $a$ is forbidden to take in state $s$; $A$ and $S$ are the action space and state space of the subordinates respectively.

A set $D$ of $suggestions$ specifies the recommendation degrees for different state-action pairs, which can be formally represented as follows,
\begin{equation} \label{eq2:suggestion}
D = \lbrace \langle s, a, d(s,a)\rangle \vert a \in A, s\in S \rbrace
\end{equation}
where $d(s,a)$ is the recommendation degree of action $a$ under state $s$. Given an action and a state $\langle s, a\rangle$, if $d(s, a) < 0$, it indicates that action $a$ is not recommended for selection in state $s$; if $d(s, a) > 0$, it indicates subordinate agents are encouraged to select action $a$ when they are in state $s$. The way of determining rules and suggestions will be covered in detail in Section \ref{section3.3.2}.

\begin{equation} \label{eq7:freq}
freq(s,a) = \frac
{ \vert \lbrace \langle s_k, a_k, r_k\rangle \mid \langle s_k, a_k, r_k\rangle \in RepInf, s_k=s, a_k=a, r_k=r_{max}(s,a) \rbrace \vert}{\vert\lbrace \langle s_k, a_k, r_k\rangle\mid \langle s_k, a_k, r_k\rangle \in RepInf, s_k=s, a_k=a \rbrace \vert}
\end{equation}

\subsection{Learning Strategy} \label{section3.3}
In this section, we first present the learning strategy of supervisors and how the rules and suggestions are generated in Section \ref{section3.3.1} and \ref{section3.3.2} respectively. Following that, we describe the learning strategy of subordinate agents and how they utilize the instructions from supervisors in Section \ref{section3.3.3}. Without loss of generality, let us assume that there is a set $S$ of supervisors, and each supervisor $i \in S$, it supervises the set $sub(i)$ of subordinate agents. Each subordinate agent $j$ has a set $neigh(j)$ of neighbors, and each supervisor agent $i$ communicates with a set $com(i)$ of other supervisors. 
\subsubsection{Supervisor's Strategy} \label{section3.3.1}
We propose that each supervisor $i$ holds a $Q$-value $Q_{i}(s,a)$ for each action $a$ under each state $s$ (row or column player). Let us denote the set of information from its subordinates as $RepInf_i= \{\langle s_k, a_k, r_k\rangle \mid k\in sub(i)\}$. For each piece of information $\langle s, a, r\rangle \in RepInf_i$, supervisor agent $i$ updates its $Q$-value following the optimistic assumption shown in Equation (\ref{eq3:q-value}),
\begin{equation} \label{eq3:q-value}
Q_{i}(s,a) = (1 - \alpha_{i}) * Q_{i}(s,a) + \alpha_{i} * r
\end{equation}
where $\alpha_{i}$ is its learning rate reflecting its updating degree between using the past experience and using the current round information.


After that, supervisor $i$ further updates its Q-values based on optimistic assumption and the frequency information of each action similar to the FMQ heuristic \cite{kapetanakis2005reinforcement}. Formally we have,
\begin{equation} \label{eq4:fmq}
FMQ_{i}(s,a) = Q_{i}(s,a) + freq(s,a) * r_{max}(s,a) * C
\end{equation}\\
where $r_{max}(s,a)$ is the max reward of each action $a$, $freq(s,a)$ is the frequency of receiving the reward of $r_{max}(s,a)$ by choosing action $a$ under state $s$ and $C$ is a weighting factor. 

The value of $r_{max}(s,a)$ is obtained from the reported information of its subordinate agents $sub(i)$. Specifically, The value of $r_{max}(s,a)$ is computed as the maximum reward that all of its subordinates receives under state $s$ by choosing action $a$ in the current round experience. Formally we have,

\begin{equation} \label{eq5:r}
\mathcal{R}(s,a) = \{r_k\mid \langle s, a, r_k\rangle \in RefInf\} 
\end{equation}
\begin{equation} \label{eq6:r max}
r_{max}(s,a) = max\{\mathcal{R}(s,a)\} 
\end{equation}

The frequency information $freq(s,a)$ is calculated as the empirical probability of receiving the maximum reward $r_{max}(s,a)$ under state $s$ when action $a$ is selected based on the reported information $RepInf$ collected from the subordinates which is shown in Equation (\ref{eq7:freq}).

After updating the strategy based on the information collected from its subordinates, we also allow each supervisor to learn from its neighboring peers (supervisors). Specifically, each supervisor communicates with a neighboring supervisor selected based on the neighboring degree between them and imitates the neighbor's strategy. The neighboring degree between a supervisor and its neighbor is related to the correlation of two groups of subordinates they supervise. A higher neighboring degree means the greater probability of choosing this neighboring supervisor. The motivation of imitating peers comes from evolutionary game theory \cite{weibull1997evolutionary}, which provides a powerful methodology to model how strategies evolve over time based on their relative performance. One of the widely used imitation rules is the proportional imitation \cite{pacheco2006coevolution}, which is adopted here as shown in Equation (\ref{eq8:p}),
\begin{equation} \label{eq8:p}
p = \frac{1}{1 + e^{-\beta * (FMQ_{j}(s,a)-FMQ_{i}(s,a))}}
\end{equation}
where the parameter $\beta$ controls the each supervisor $i$'s degree of imitating the strategy (the $FMQ$-value) of the neighboring supervisor $j$.

Finally each supervisor $i$ updates its strategy (denoted as $E$-value $E_{i}(s,a)$) for each action $a$ under state $s$ as the average between the FMQ-values of its own and its neighbor $j$ weighted by parameter $p$. Formally we have,
\begin{equation} \label{eq9:e-value}
E_{i}(s,a) = (1 - p) * FMQ_{i}(s,a) + p * FMQ_{j}(s,a)
\end{equation}

\subsubsection{Supervisor Instruction Generation} \label{section3.3.2}
Next we introduce how a supervisor generates instructions for its subordinates at the end of each round. As previously mentioned, there are two forms of instructions from a supervisor: rules and suggestions. First each supervisor $i$ normalizes the $E$-values, which serves as the basis for generating instructions for its subordinates. Formally we have,
\begin{equation} \label{eq10:d}
E^\prime_i(s,a) = \frac{E_i(s,a) - \overline{E_i(s)}}{\sigma}
\end{equation}
where $\overline{E_i(s)}$ is the mean of $E$-values averaged over all action estimates in state $s$ and as shown in Equation (\ref{eq11:mean}),
\begin{equation} \label{eq11:mean}
\overline{E_i(s)} = \frac{\sum_{a\in A} E_i(s,a)}{|A|}
\end{equation}
The parameter $\sigma$ is the standard deviation of $FMQ$-value following Equation (\ref{eq12:sd}),
\begin{equation} \label{eq12:sd}
\sigma = \sqrt{\frac{1}{|A|}\sum_{a\in A}(E_i(s,a)- \overline{E_i(s,a)})^{2}}
\end{equation}

Given a state-action pair $\langle s, a \rangle$, if the $E^\prime$-value $E^\prime(s,a)$ is smaller than a given threshold, it indicates that selecting action $a$ is not a wise choice under state $s$, thus it is encoded as a rule. Formally we have,

\begin{equation} \label{eq13:f}
F = \lbrace \langle s, a\rangle \vert E^\prime(s,a) < \delta  \rbrace
\end{equation}
where $\delta$ is the threshold which is set to the value of -0.5 in this paper.

For each state-action pair $(s,a)$, its recommendation degree $d(s,a)$ is set to the value of $E^\prime_i(s,a)$. Thus the set of suggestions from supervisor $i$ can be represented as follows,
\begin{equation}
D = \{\langle s, a, E^\prime_i(s, a)\rangle\mid a\in A, s\in S\}
\end{equation}
Given a state-action pair $(s,a)$, if $E^\prime_i(s, a) < 0$, it indicates that selecting action $a$ is not recommended under state $s$; if $E^\prime_i(s, a) > 0$, it indicates subordinate agents are encouraged to select action $a$ when they are in state $s$.

\subsubsection{Learning Strategy of Subordinates} \label{section3.3.3}
Similar to the strategies of supervisors, each subordinate agent $j$ also keeps a record of a $Q$-value $Q_{j}(s,a)$ for each action $a \in A_j$ under each state $s$. The Q-value $Q_{j}(s,a)$ indicates the past performance of choosing action $a$ under state $s$ and serves as the basis for making decisions [3]. For each subordinate agent $j$, let us first denote its feedback information received by the end of round $t$ as $FeedInf^t_{j} = \lbrace \langle s_m, a_m, r_m \rangle \vert m \in [1,t]\}$. At the end of each round $t$, subordinate agent $j$ updates its $Q$-value based on its feedback $\langle s_t, a_t, r_t\rangle$ as follows,
\begin{equation} \label{eq14:q-value}
Q_{j}(s_t,a_t) = (1 - \alpha_{j}) * Q_{j}(s_t,a_t) + \alpha_{j} * r_t
\end{equation}
where $\alpha_{j}$ is the learning rate modelling its updating degree between using the previous experience and using the most recent information.

Additionally, each subordinate agent also updates its Q-values by taking into consideration both the optimistic assumption and the frequency information \cite{kapetanakis2005reinforcement}. Formally we have,
\begin{equation} \label{eq15:fmq-value}
FMQ_{j}(s,a) = Q_{j}(s,a) + freq(s,a) * r_{max}(s,a) * C
\end{equation}
where $r_{max}(s,a)$ is the max reward of each action $a$ based on its own experience, $freq(s,a)$ is the frequency of getting the payoff of $r_{max}(s,a)$ until now for action $a$ and $C$ is a weighting factor defining the trade-off between updating using Q-values and maximum payoff information. $freq(s,a)$ is calculated the same as shown in Equation (\ref{eq7:freq}).

After receiving supervisor's suggestions, each subordinate further adjusts its estimation of the goodness of each state-action pair based on the FMQ-values as follows,

\begin{equation} \label{eq17:e-value}
E_{j}(s,a) = FMQ_{j}(s,a) * (1 + d(s,a)*\rho)
\end{equation}
where $d(s,a)$ is the suggestion degree on the state-action pair $(s, a)$, and $\rho$ is a weighting factor controlling the influence of the recommendation degree on the E-values.

Besides, supervisors also influence the subordinate agents' exploration rates. Let us suppose a subordinate agent $j$ selects action $a$ under current state $s$. If the supervisor $i$'s recommendation degree $d(s,a) < 0$, which indicates subordinate agent $j$'s current choice is not recommended, and agent $j$ should increase the exploration rate to have more chance to select the recommended actions next time. On the other hand, if the recommendation degree $d(s, a) > 0$, it indicates the subordinates' current choice is recommended. Thus subordinate agent $j$ decreases its exploration rate to avoid selecting discouraging actions in the future. For both cases, the adjustment degree varies depending on the absolute value of the state-action pair's recommendation degree. Formally each subordinate agent updates its exploration rate as follows,
\begin{equation}
\label{eq18:epsilon}
    \epsilon_j = \epsilon_j * (1 - d(s,a)*\gamma) 
\end{equation}
where $\gamma$ is a weighting factor controlling the influence degree of the supervisor's suggestion on the subordinates' exploration rates.

Finally, given the current state $s$, each subordinate agent $j$ chooses its action from those actions whose corresponding state-action pair do not belong to the set $F$ of rules based on the corresponding set of E-values according to the $\epsilon$-greedy mechanism. The probability $\pi_{j}(s,a)$ for choosing action $a$ under state $s$ is defined in Equation \ref{eq19:strategy}. Specifically, each agent chooses its action with the highest E-value with probability $1-\epsilon_j$ to exploit the action with best performance currently (random selection in case of a tie), and makes random choices with probability $\epsilon_j$ to explore new actions with potentially better performance. 
\begin{equation}
\label{eq19:strategy}
\pi_j(s,a) =
 \left\{
  \begin{array}{ll}
     1 - \epsilon_j & \quad \text{if $a$ = $argmax_{a^{'}}E(s,a^{'})$}\\
     0 & \quad \text{if $a \in F_{j}$} \\
     \frac{\epsilon_j}{|A_j|-|F_{j}| - 1}  & \quad \text{otherwise}      
  \end{array} \right.
\end{equation}

\section{Experimental Simulation} \label{section4}
In this section, we start with evaluating the norm emergence performance of our approach HHLS under different types of games by comparing with the state-of-the-art strategies. Following that we explore the influence of some key parameters on norm emergence. We use degree grouping in the following experimental simulations, and the influence of different grouping mechanisms are investigated in Section \ref{section4.1}. All results are averaged over 1000 runs. The parameter settings and explanations are shown in Table \ref{table4} and Table \ref{tableex} respectively.
\begin{table}
\begin{center}
{\caption{The initial value of parameters.}\label{table4}}
\begin{tabular}{ccccccc}
\hline
\rule{0pt}{12pt}
Parameters & $\alpha$ & $\epsilon$ & $\beta$ & $\gamma$ & $\rho$ & $d$ \\
\hline
\rule{0pt}{12pt}
Value & 0.99 & 0.93 & 0.1 & 0.05 & 0.01 & 6 \\
\hline
\end{tabular}
\end{center}
\end{table}

\begin{table}
\begin{center}
{\caption{The explanations of parameters.}\label{tableex}}
\begin{tabular}{ccccccc}
\hline
\rule{0pt}{12pt}
Parameters & Explanations \\
\hline
\rule{0pt}{12pt}
$\alpha$ & Learning rate \\
\hline
\rule{0pt}{12pt}
$\epsilon$ & $\epsilon$-greedy probability \\
\hline
\rule{0pt}{12pt}
$\beta$ & Weighting factor on supervisors' interaction \\
\hline
\rule{0pt}{12pt}
$\gamma$ & Weighting factor on subordinates' exploration rates\\
\hline
\rule{0pt}{12pt} 
$\rho$ & Weighting factor on subordinates' E-Value \\
\hline
\rule{0pt}{12pt} 
$d$ & Average connection degree of small-world and scale-free network \\
\hline
\end{tabular}
\end{center}
\end{table}

\subsection{Performance Evaluation in Different Games} \label{section4.2}
We compare our approach HHLS with two previous works: hierarchical learning in \cite{yu2015hierarchical} and social learning in \cite{airiau2014emergence}. All these three learning approaches are within the same social learning environment, i.e., each agent is allowed to interact with only one of its neighbors each round. The work in \cite{airiau2014emergence} is the representative state-of-the-art approach tackling norm emergence problem under multiagent social learning framework without considering any hierarchical organization. The work in \cite{yu2015hierarchical} is the most recent approach introducing hierarchical learning into multiagent social learning framework to improve norm emergence efficiency. Four representative 6-action games are considered shown from Table \ref{table5} to Table \ref{table8}. 

\subsubsection{Coordination Game (CG)} \label{section4.2.1}
\begin{figure*}
   \centering
   \begin{minipage}[t]{.48\linewidth}    
	\centerline{\includegraphics[height = 2in, width = 2.4in]{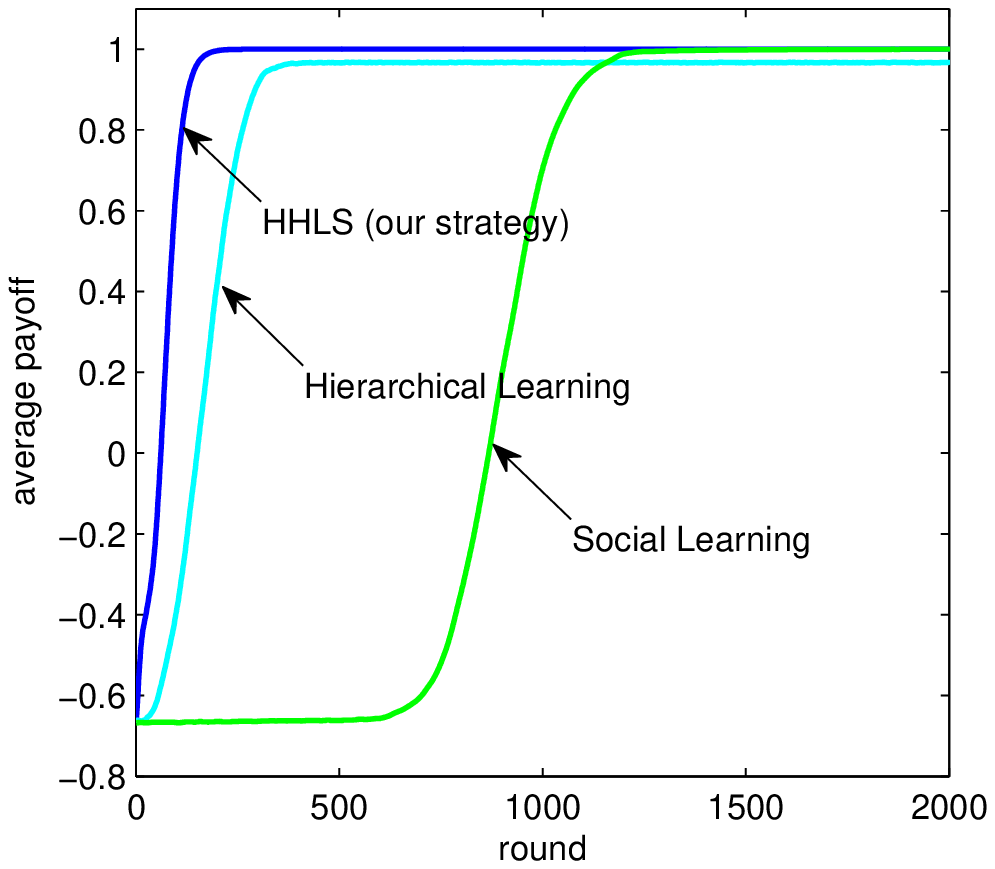}}
	\caption{The dynamics of the average payoffs of agents in coordination games under different strategies.} \label{figure2}
  \end{minipage}
  \hfill
  \begin{minipage}[t]{.48\linewidth}   
  	\centerline{\includegraphics[height = 2in, width = 2.4in]{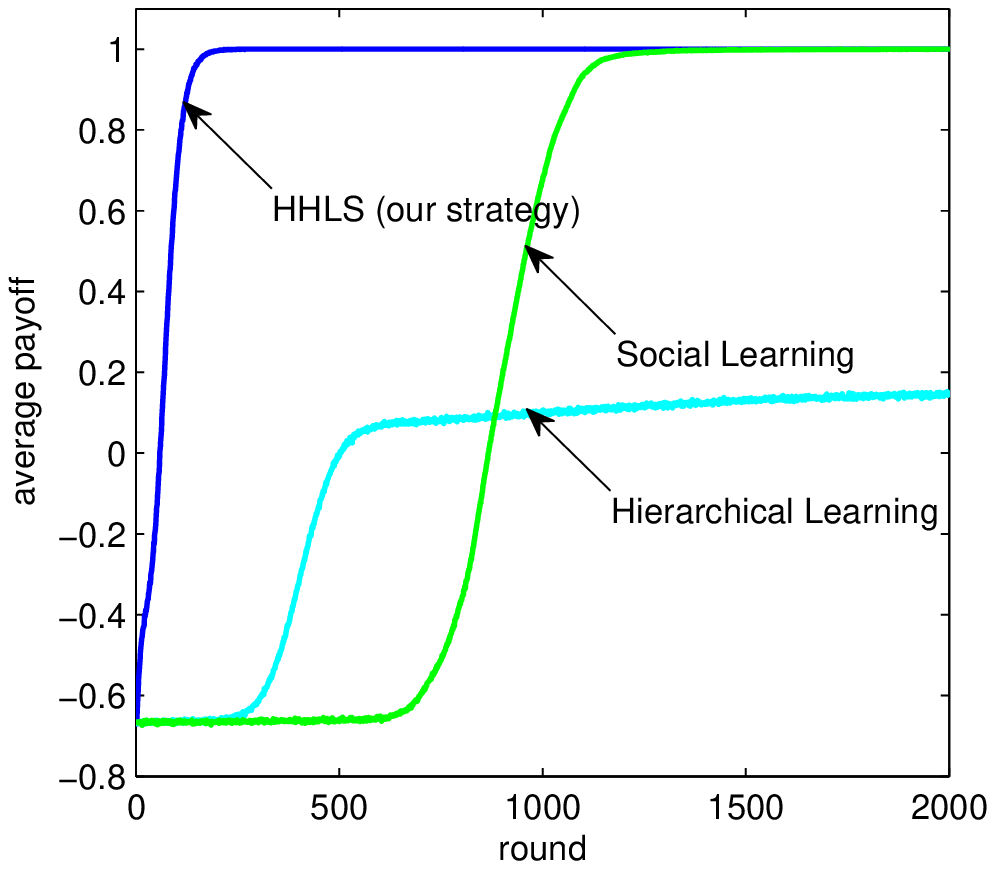}}
  	\caption{The dynamics of the average payoffs of agents in anti-coordination games under different strategies.} \label{figure3}
  \end{minipage}
  \vfill
\end{figure*}

We first consider agents playing a 6-action coordination game (Table \ref{table5}) in which there exist six norms marked with red. Agents are preferred to choose the same action. Fig \ref{figure2} shows the dynamics of the average payoffs of agents with the number of rounds averaged for the three learning approaches. We can observe that all learning methods enable agents to achieve an average payoff of 1. Our hierarchically heuristic learning strategy converges faster than the hierarchical learning method \cite{yu2015hierarchical}, and the social learning method \cite{airiau2014emergence} is the slowest. This is because HHLS enables the supervisor agents to influence subordinate agents in a more efficient manner, thus accelerating norm emergence.

\begin{table}  
\caption{The payoff matrix of coordination game.}\label{table5}
\centering
\begin{tabular}{cccccccc}
\hline 
\rule{0pt}{12pt}
\multirow{2}{0.5 in}{} &  & \multicolumn{6}{c}{Agent 2's actions} \\ 
\cline{3-8}
\rule{0pt}{12pt}
& & a & b & c & d & e & f \\ 
\hline
\multirow{6}{0.5 in}{Agent 1's\\ actions}\\
 & a & {\color{red}1} & -1 & -1 & -1 & -1 & -1 \\ 
\cline{2-8}
\rule{0pt}{12pt}
 & b & -1 & {\color{red}1} & -1 & -1 & -1 & -1 \\ 
\cline{2-8}
\rule{0pt}{12pt}
 & c & -1 & -1 & {\color{red}1} & -1 & -1 & -1 \\ 
\cline{2-8}
\rule{0pt}{12pt}
 & d & -1 & -1 & -1 & {\color{red}1} & -1 & -1 \\ 
\cline{2-8}
\rule{0pt}{12pt}
 & e & -1 & -1 & -1 & -1 & {\color{red}1} & -1 \\ 
\cline{2-8}
\rule{0pt}{12pt}
 & f & -1 & -1 & -1 & -1 & -1 & {\color{red}1} \\ 
\hline 
\end{tabular}
\end{table}

\begin{table}
\caption{The payoff matrix of anti-coordination game.}\label{table6}
\centering
\begin{tabular}{cccccccc}
\hline 
\rule{0pt}{12pt}
\multirow{2}{0.5 in}{} &  & \multicolumn{6}{c}{Agent 2's actions} \\ 
\cline{3-8}
\rule{0pt}{12pt}
& & a & b & c & d & e & f \\ 
\hline
\multirow{6}{0.5 in}{Agent 1's\\ actions}\\
 &a & -1 & -1 & -1 & -1 & -1 & {\color{red}1} \\ 
\cline{2-8}
\rule{0pt}{12pt}
 &b & -1 & -1 & -1 & -1 & {\color{red}1} & -1 \\ 
\cline{2-8}
\rule{0pt}{12pt}
 &c & -1 & -1 & -1 & {\color{red}1} & -1 & -1 \\ 
\cline{2-8}
\rule{0pt}{12pt}
 &d & -1 & -1 & {\color{red}1} & -1 & -1 & -1 \\ 
\cline{2-8}
\rule{0pt}{12pt}
 &e & -1 & {\color{red}1} & -1 & -1 & -1 & -1 \\ 
\cline{2-8}
\rule{0pt}{12pt}
 &f & {\color{red}1} & -1 & -1 & -1 & -1 & -1 \\ 
\hline 
\end{tabular}
\end{table}
\subsubsection{Anti-coordination Game (ACG)} \label{section4.2.2}
Similarly, we consider agents playing a 6-action anti-coordination game (Table \ref{table6}) in which there also exist six equivalently optimal norms marked with red. However, different from coordination game, each norm requires agents to choose different actions. Fig \ref{figure3} shows the dynamics of the average payoffs using three learning methods. We can observe that both social learning \cite{airiau2014emergence} and HHLS enable agents to achieve an average payoff of 1, while the hierarchical learning fails. Besides, our HHLS converge faster than the social learning approach \cite{airiau2014emergence}, which justifies the efficiency of introducing a hierarchical learning structure. For the hierarchical learning \cite{yu2015hierarchical}, it does not distinguish the state information and thus cannot adaptively select different actions for different states. 

\begin{table}
\centering
\caption{The payoff matrix of coordination game with high penalty.}\label{table7}
\begin{tabular}{cccccccc}
\hline 
\rule{0pt}{12pt}
\multirow{2}{0.5 in}{}&  & \multicolumn{6}{c}{Agent 2's actions} \\ 
\cline{3-8}
\rule{0pt}{12pt}
& & a & b & c & d & e & f \\ 
\hline
\multirow{6}{0.5 in}{Agent 1's\\ actions}\\
 &a & {\color{red}10} & 0 & -30 & -30 & 0 & -30 \\ 
\cline{2-8}
\rule{0pt}{12pt}
 &b & 0 & {\color{blue}7} & 0 & 0 & 0 & 0 \\ 
\cline{2-8}
\rule{0pt}{12pt}
 &c & -30 & 0 & {\color{red}10} & -30 & 0 & -30 \\ 
\cline{2-8}
\rule{0pt}{12pt}
 &d & -30 & 0 & -30 & {\color{red}10} & 0 & -30 \\ 
\cline{2-8}
\rule{0pt}{12pt}
 &e & 0 & 0 & 0 & 0 & {\color{blue}7} & 0 \\ 
\cline{2-8}
\rule{0pt}{12pt}
 &f & -30 & 0 &-30 & -30 & 0 & {\color{red}10} \\ 
\hline 
\end{tabular}
\end{table}
\subsubsection{Coordination Game with High Penalty (CGHP)} \label{section4.2.3}
\begin{figure*}
  \centering
  \begin{minipage}[t]{.48\linewidth}
  	\centerline{\includegraphics[height = 2in, width = 2.4in]{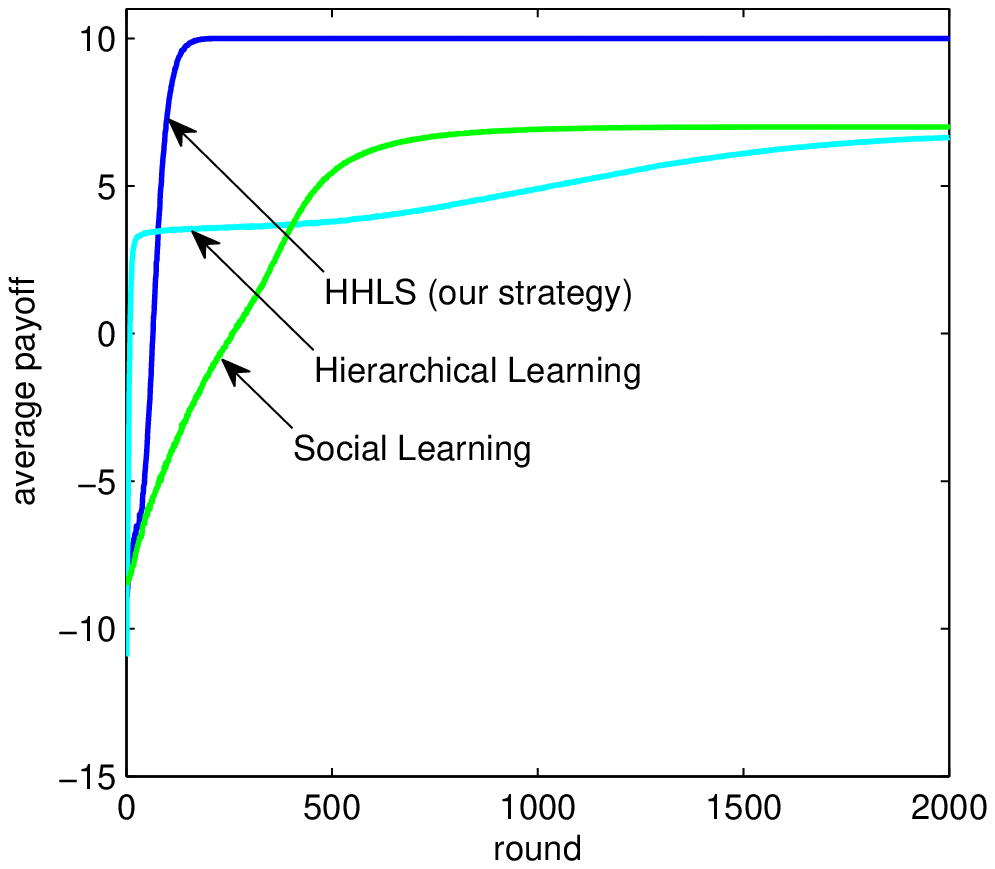}}
	\caption{The dynamics of the average payoffs of agents in CGHP under different strategies.} \label{figure4}
  \end{minipage}
  \hfill
  \begin{minipage}[t]{.48\linewidth}    
	\centerline{\includegraphics[height = 2in, width = 2.4in]{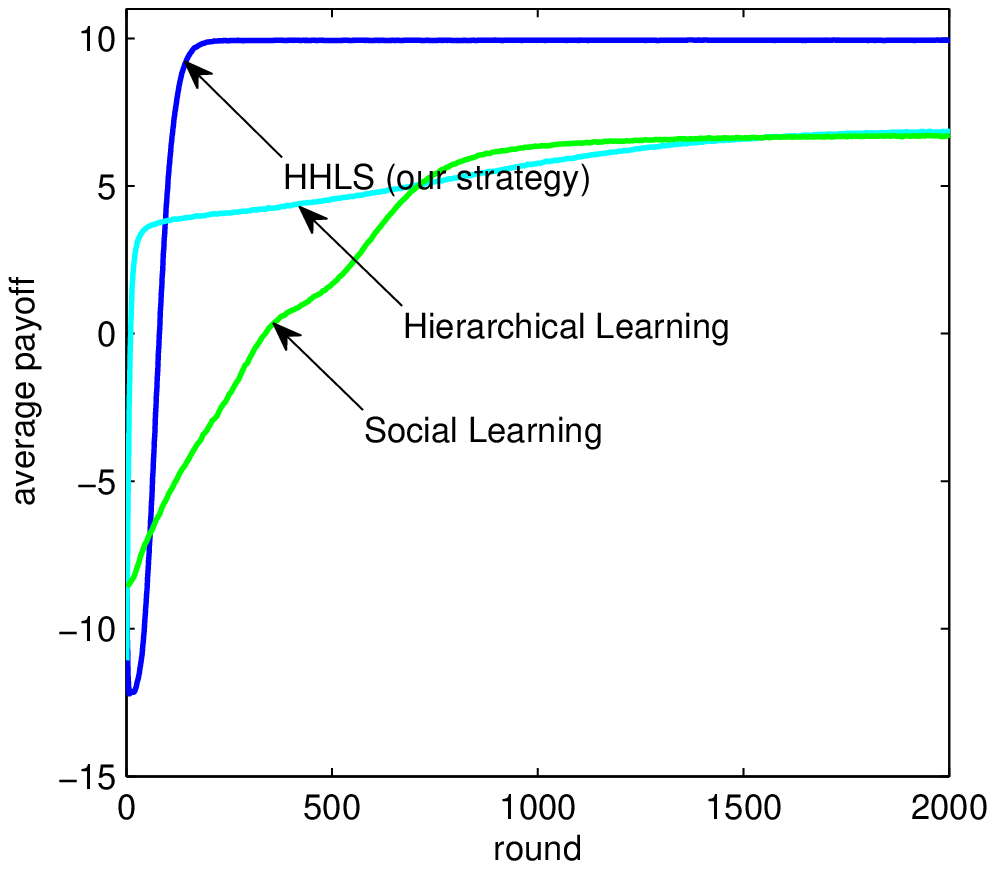}}
	\caption{The dynamics of the average payoffs of agents in FSCGHP under different strategies.} \label{figure5}
  \end{minipage}
  \vfill
  \end{figure*}

Next, we consider 100 agents playing a 6-action coordination game with high penalty (Table \ref{table7}), in which there exist four optimal norms marked with red and two suboptimal norms marked with blue. In this kind of games, agents are vulnerable to converge to suboptimal norms due to the existence of high mis-coordination cost (-30). Fig \ref{figure4} shows the dynamics of the average payoffs of agents with the number of rounds for the three learning approaches. We can see that only HHLS enables agents to achieve an average payoff of 10 (i.e., converging to one optimal norm). The other two learning methods converge to one of the suboptimal norms, and they also converge slower than HHLS. We hypothesize the superior performance of HHLS is due to the integration of optimistic assumption during strategy update (to overcome mis-coordination cost effect) and efficient hierarchical supervision (to accelerate norm emergence speed). 

\subsubsection{Fully Stochastic Coordination Game with High Penalty (FSCGHP)} \label{section4.2.4}
Last, we consider agents playing a 6-action fully stochastic coordination game with high penalty (Table \ref{table8}). In FSCGHP, each outcome is associated with two possible payoffs and the agents receive one of them with probability 0.5, which models the uncertainty of the interaction results. The expected payoffs of this game are the same with the CGHP and there also exist four optimal norms marked with red and two suboptimal norms marked with blue. But it is more complex and difficult to emerge norms due to the stochasticity of the environments. Fig \ref{figure5} shows the dynamics of the average payoffs of agents against the number of rounds for the three learning strategies. We can observe that in this challenging game, only HHLS enables agents to achieve an average payoff of 10 (agents learn to converge to one optimal). In contrast, the other two learning strategies converge to one of the suboptimal norms with a slower convergence rate. Finally, it is worth mentioning that if the size of the action, and norm space is further increased, the social learning method \cite{airiau2014emergence} and hierarchical learning method \cite{yu2015hierarchical} cannot converge (to a suboptimal norm) within 10000 runs. However, HHLS still can support converging to one optimal norm within approximately 200 rounds. The influence of action size will be discussed in more detail in Section \ref{section4.4.2}.
\begin{table}
\centering
\caption{The payoff matrix of fully stochastic coordination game with high penalty.}\label{table8}
\begin{tabular}{cccccccc}
\hline 
\rule{0pt}{12pt}
\multirow{2}{0.5 in}{1's payoff \\ 2's payoff}&  & \multicolumn{6}{c}{Agent 2's actions} \\ 
\cline{3-8}
\rule{0pt}{12pt}
& & a & b & c & d & e & f \\ 
\hline
\multirow{6}{0.5 in}{Agent 1's \\actions}\\
 & a & {\color{red}12/8} & 5/-5 & -20/-40 & -20/-40 & 5/-5 & -20/-40 \\ 
\cline{2-8}
\rule{0pt}{12pt}
 & b & 5/-5 & {\color{blue}14/0} & 5/-5 & 5/-5 & 5/-5 & 5/-5 \\ 
\cline{2-8}
\rule{0pt}{12pt}
 & c & -20/-40 & 5/-5 & {\color{red}12/8} & -20/-40 & 5/-5 & -20/-40 \\ 
\cline{2-8}
\rule{0pt}{12pt}
 & d & -20/-40 & 5/-5 & -20/-40 & {\color{red}12/8} & 5/-5 & -20/-40 \\ 
\cline{2-8}
\rule{0pt}{12pt}
 & e & 5/-5 & 5/-5 & 5/-5 & 5/-5 & {\color{blue}14/0} & 5/-5 \\ 
\cline{2-8}
\rule{0pt}{12pt}
 & f & -20/-40 & 5/-5 & -20/-40 & -20/-40 & 5/-5 & {\color{red}12/8} \\ 
\hline 
\end{tabular}
\end{table}

\subsection{Influence of Different Components of HHLS} \label{section4.3}
Next, to justify the importance of the key components (optimistic assumption and supervisor's guidance) in our HHLS framework, we further investigate the relative importance of different modified versions of HHLS mechanisms: HHLS, HHLS without optimistic assumption (i.e., HHLS without FMQ heuristic), HHLS without Supervisors' guidance (i.e., HHLS without hierarchical structure).  We use coordination game with high penalty (CGHP) as an illustrating example and the results are similar under other matrix games.

\begin{figure*}
   \centering
  \hfill
  \begin{minipage}[t]{.46\linewidth}   
  	\centerline{\includegraphics[height = 2in, width = 2.4in]{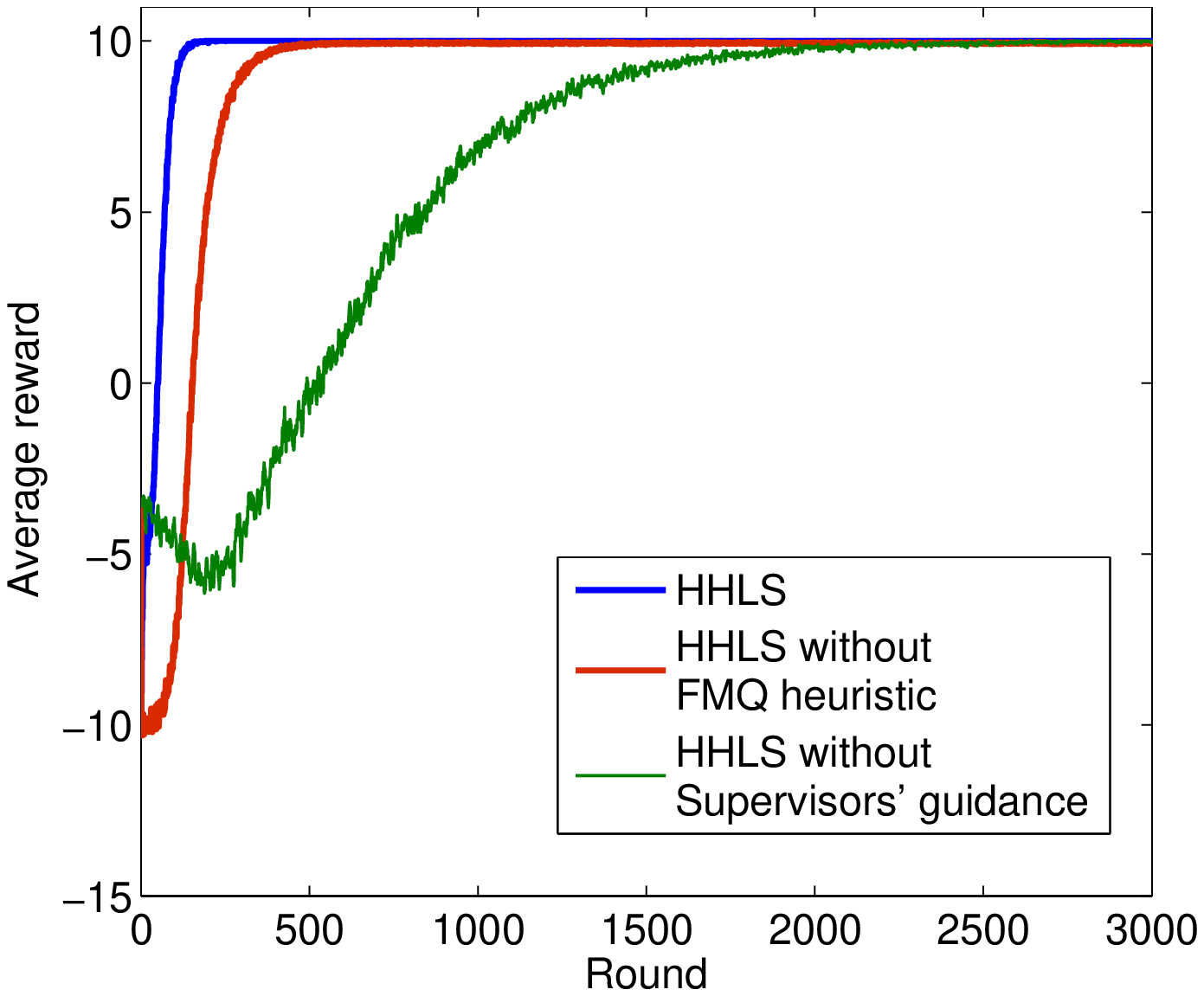}}
  	\caption{The dynamics of the average payoffs of agents in 3-action CGHP under different HHLS mechanisms.} \label{figure20}
  \end{minipage}
  \hfill
  \begin{minipage}[t]{.46\linewidth}
  	\centerline{\includegraphics[height = 2in, width = 2.4in]{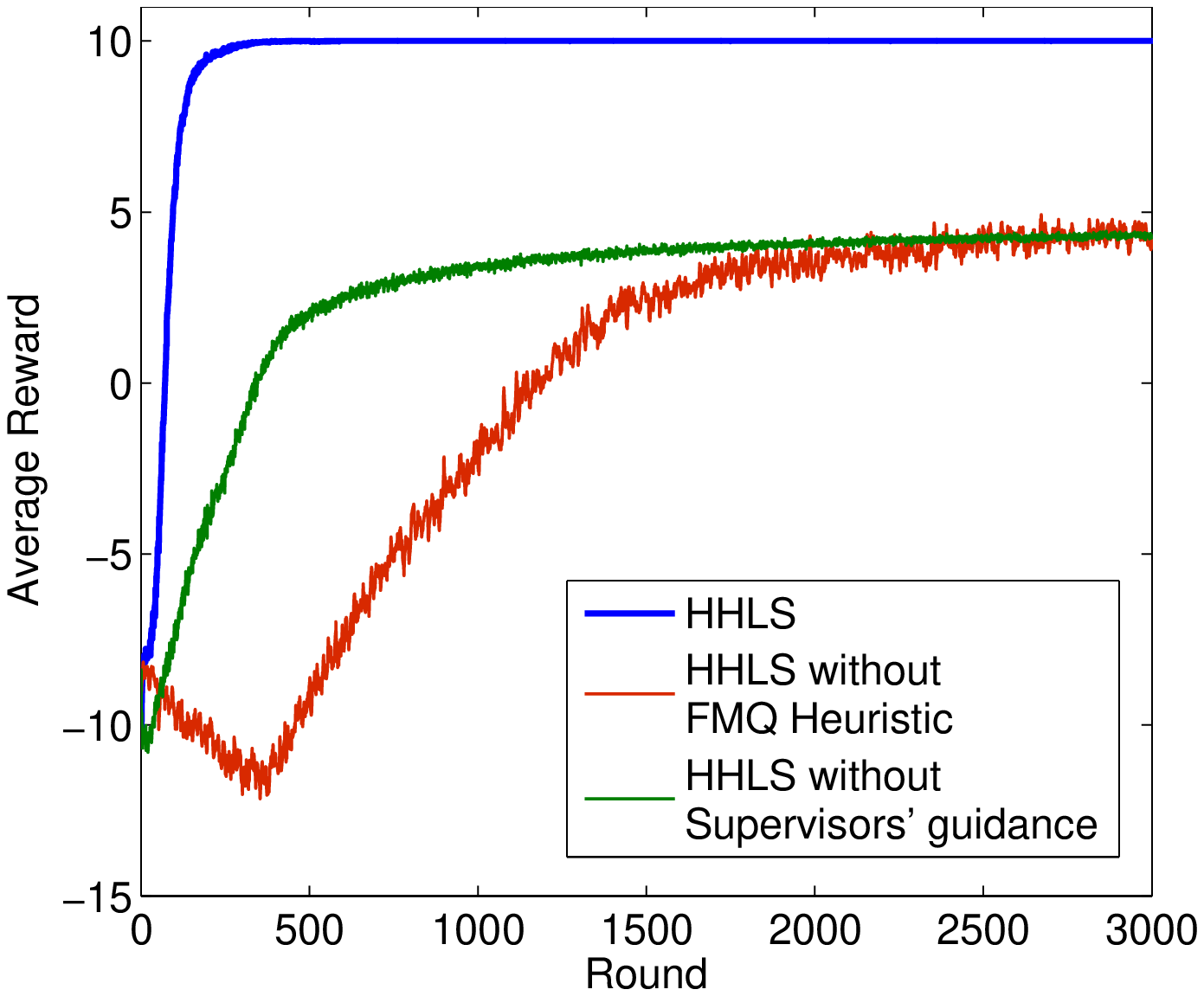}}
	\caption{The dynamics of the average payoffs of agents in 6-action CGHP under different strategies.} \label{figure21}
  \end{minipage}
  \vfill
\end{figure*}

Fig \ref{figure20} shows the different average payoffs under three different HHLS mechanisms for a small-size CGHP with 3 actions. We can see that HHLS without optimistic assumption or without supervisors' guidance still enables agents to converge to an optimal norm. However, the convergence rate under these two mechanisms is much slower than the full HHLS mechanism. Next, we consider increasing the game size to 6 actions and the results are given in Fig \ref{figure21}. We can see that only the full HHLS mechanism can enable fast convergence to the optimal norm while the other two approaches fail. These results confirm the hypothesis proposed in Section \ref{section4.2.3} that both components are critical and the integration of the two components leads to superior performance of HHLS. 

\subsection{Performance of Different Grouping Mechanisms} \label{section4.1}
In this section, we investigate the performance of norm emergence using different grouping mechanisms described above in Section \ref{section3.4}.

\begin{figure*}
 \centering
 \begin{minipage}[t]{.5\linewidth}  
 \centerline{\includegraphics[height = 2in, width = 2.5in]{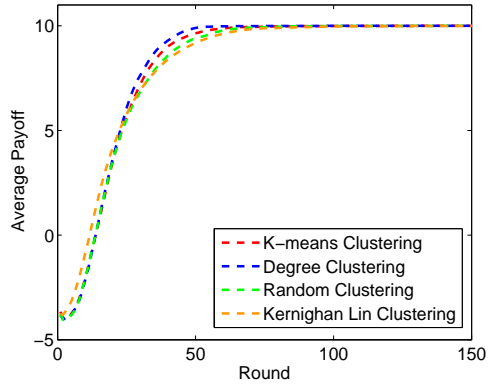}}
  	\caption{The influence of different grouping mechanisms on the performance of norm emergence.} \label{figure18}
  	\end{minipage}
\hfill
  \vfill
  \end{figure*}

Fig \ref{figure18} shows the influence of different grouping mechanisms on the performance of norm emergence. The results are averaged over 1000 iterations in 3-action coordination game with high penalty with a population of 100 agents. We can see that all these grouping mechanisms enable efficient norm emergence while degree grouping outperforms the rest of three grouping mechanisms. The slight differences among these grouping mechanisms prove to be statistically significant under t-test.

Intuitively, those subordinate agents with higher degrees play a more important role in the whole population because they can influence more neighboring agents. Degree grouping allocates agents with the same degree into the same group. For more important agents (with higher degree), they can reach a unified opinion more quickly, and in turn influence the rest of agents of lower degrees to quicker convergence. 

\subsection{Influence of Non-hierarchical Parameters} \label{section4.4}
In this section, we investigate the influence of key parameters on the performance of norm emergence. We present the following results for hierarchically heuristic learning using a small-world network. Except the parameter whose effect is being investigated, the rest of parameters follow the same settings as in Section \ref{section4}.

\subsubsection{Influence of Population Size} \label{section4.4.1}
\begin{figure*}
  \centering
  \begin{minipage}[t]{.48\linewidth}   
  	\centerline{\includegraphics[height = 2in, width = 2.4in]{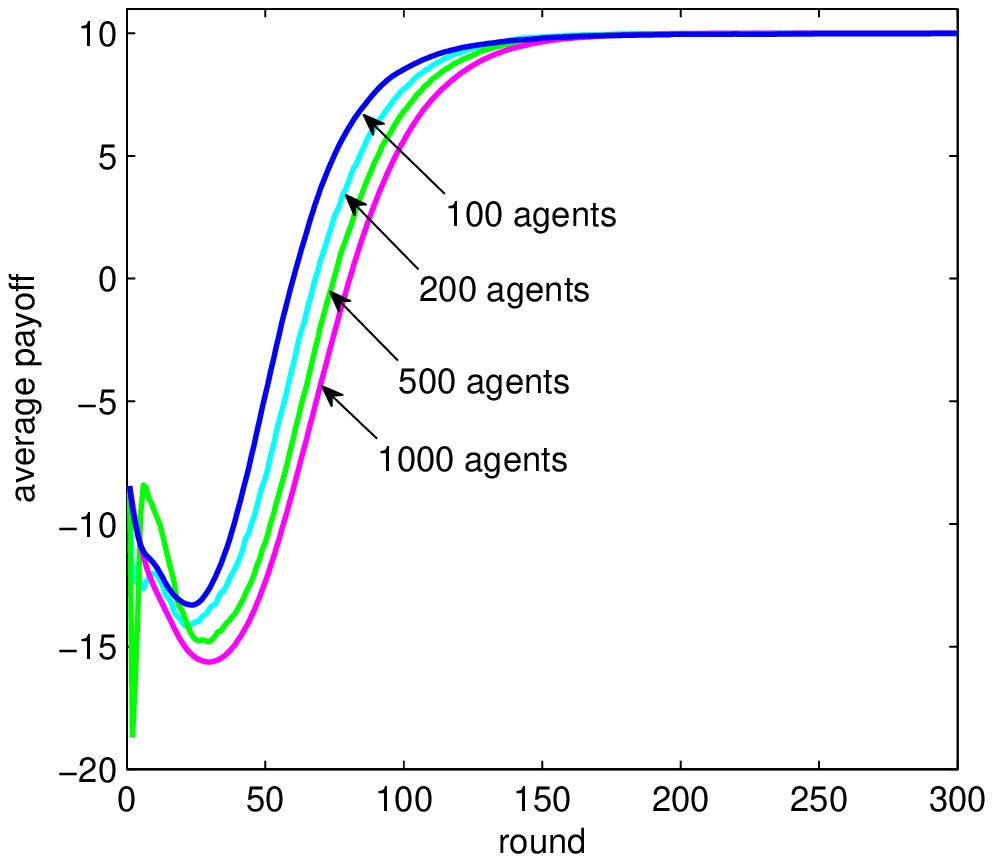}}
  	\caption{The influence of population size.} \label{figure6}
  \end{minipage}
  \hfill
  \begin{minipage}[t]{.48\linewidth}
  	\centerline{\includegraphics[height = 2in, width = 2.4in]{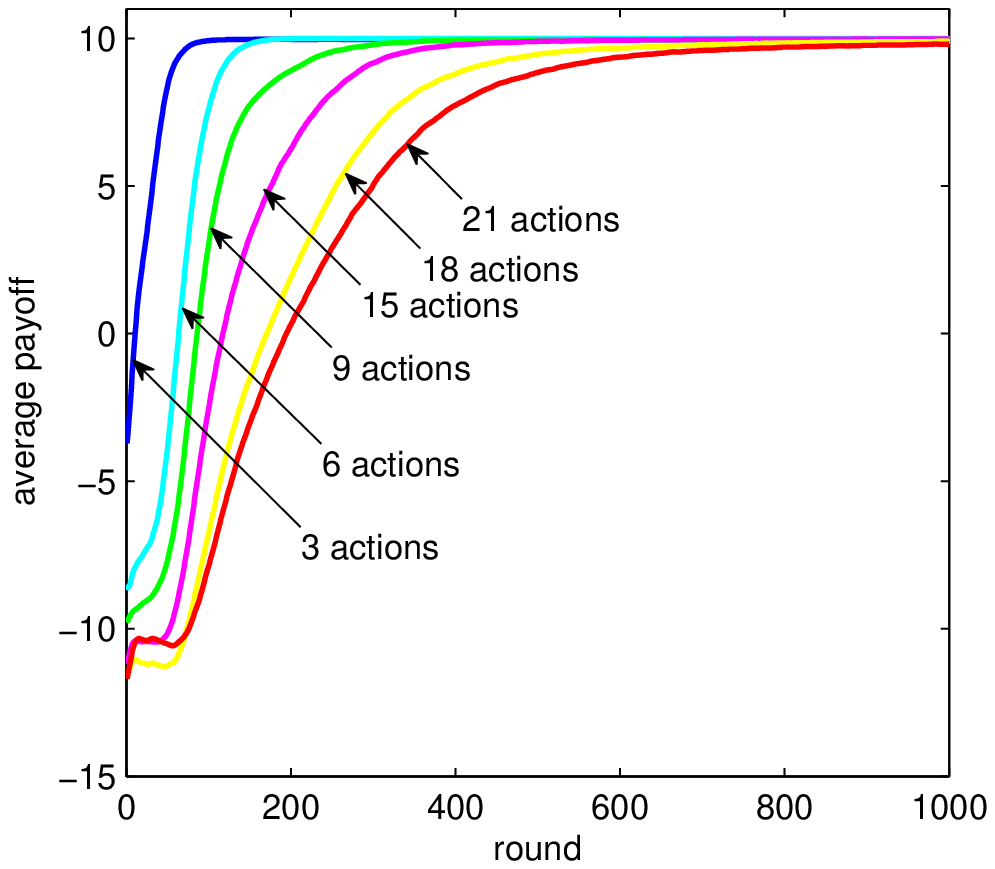}}
  	\caption{The influence of action size.} \label{figure7}
  \end{minipage}
  \vfill
  \end{figure*}

The influence of population size is shown in Fig \ref{figure6}. For a 6-action coordination game with high penalty (CGHP), we can clearly observe the norm emergence efficiency is reduced with the increase of the population size. As the size of clusters is held constant, the number of clusters increase as the population size becomes larger. We hypothesize that the increase of clusters impede efficient coordination among agents in the hierarchical framework in the following two ways. First, the coordination efforts required among cluster supervisors increase significantly with the number of clusters. Second, the increase of clusters also add difficulty to the efficient guidance of each supervisor to generate suggestions to guide all of their subordinate agents towards a consistent norm.
\subsubsection{Influence of Action Size} \label{section4.4.2}
Fig \ref{figure7} shows the dynamics of the average payoffs of agents for different action sizes on the 6-action coordination game with high penalty (CGHP). We can see that the convergence rate decreases with increase in the size of the action space. This is expected because the coordination space becomes larger when the action size increases. Besides, larger action size usually results in more chances of mis-coordination cost and suboptimal norms, which additionally increases the coordination difficulty for agents selecting a consistent norm. Finally, it is worth noting that with the increase in the action space size, our framework can still efficiently support norm emergence without significantly degrading the performance (norm emergence occurs within 1000 rounds for all cases). In contrast, in previous socially learning framework without utilizing a hierarchical organization \cite{airiau2014emergence}, the norm convergence speed is significantly reduced when the action space is increased.

\begin{table}
\begin{center}
{\caption{The average number of rounds needed before convergence under different network topologies.}\label{table9}}
\begin{tabular}{cccccc}
\hline 
\rule{0pt}{12pt}
\multirow{2}{0.2 in}{Convergence Speed}& & \multicolumn{4}{c}{Game type} \\ 
\cline{3-6}
\rule{0pt}{12pt}
& & CG & ACG & CGHP & FSCGHP \\ 
\hline
\multirow{5}{0.5 in}{Network \\ topology}\\
 & Grid & 142 & 141 & 131 & 153 \\ 
\cline{2-6}
\rule{0pt}{12pt}
 & Ring & 144 & 146 & 135 & 157 \\ 
\cline{2-6}
\rule{0pt}{12pt}
 & Random & 146 & 136 & 122 & 162 \\ 
\cline{2-6}
\rule{0pt}{12pt}
 & Small-world & 141 & 141 & 124 & 162\\ 
\cline{2-6}
\rule{0pt}{12pt}
 & Scale-free & 149 & 144 & 129 & 163 \\ 
\hline 
\end{tabular}
\end{center}
\end{table}

\subsubsection{Influence of Network Topology} \label{section4.4.3}
We evaluate the influence of five different networks: random network, grid network, ring network, small-world and scale-free network on the 6-action coordination game with high penalty (CGHP). Table \ref{table9} shows the average number of rounds needed before convergence. We find that hierarchical social learning framework performs robustly under different network topologies. HHLS enables agents to converge to norms in approximately the same number of rounds under all the above five network topologies for the different game types we have experimented with.

\begin{table}
\begin{center}
{\caption{The influence of neighborhood size.}\label{table10}}
\begin{tabular}{ccccccccc}
\hline
\rule{0pt}{12pt}
Neighbor \\ size & 2 & 6 & 8 & 10 & 20 & 30 & 50 & 99\\
\hline
\rule{0pt}{12pt}
Convergence \\Rate & 144 & 141 & 143 & 139 & 141 & 141 & 142 & 139\\
\hline
\end{tabular}
\end{center}
\end{table}

\subsubsection{Influence of Neighborhood Size} \label{section4.4.4}
We empirically evaluate the influence of neighborhood size varying it from 2 up to 99 (fully connected) with a population of 100 agents on the 6-action coordination game with high penalty (CGHP). Table \ref{table10} shows the average number of rounds needed before convergence for different neighborhood sizes. We can see that the average number of rounds required stabilizes at around 140 rounds. This finding is different from the results usually observed in the traditional socially learning framework without a hierarchical structure \cite{airiau2014emergence}. This is because in hierarchical social learning framework, each supervisor supervises and guides a cluster of subordinate agents, which can overcome the low connectivity disadvantage when the neighborhood size is small.

\subsection{Influence of Hierarchical Factors} \label{section4.5}
\begin{figure*} 
  \begin{minipage}[t]{.48\linewidth}
  	\centerline{\includegraphics[height = 2in, width = 2.4in]{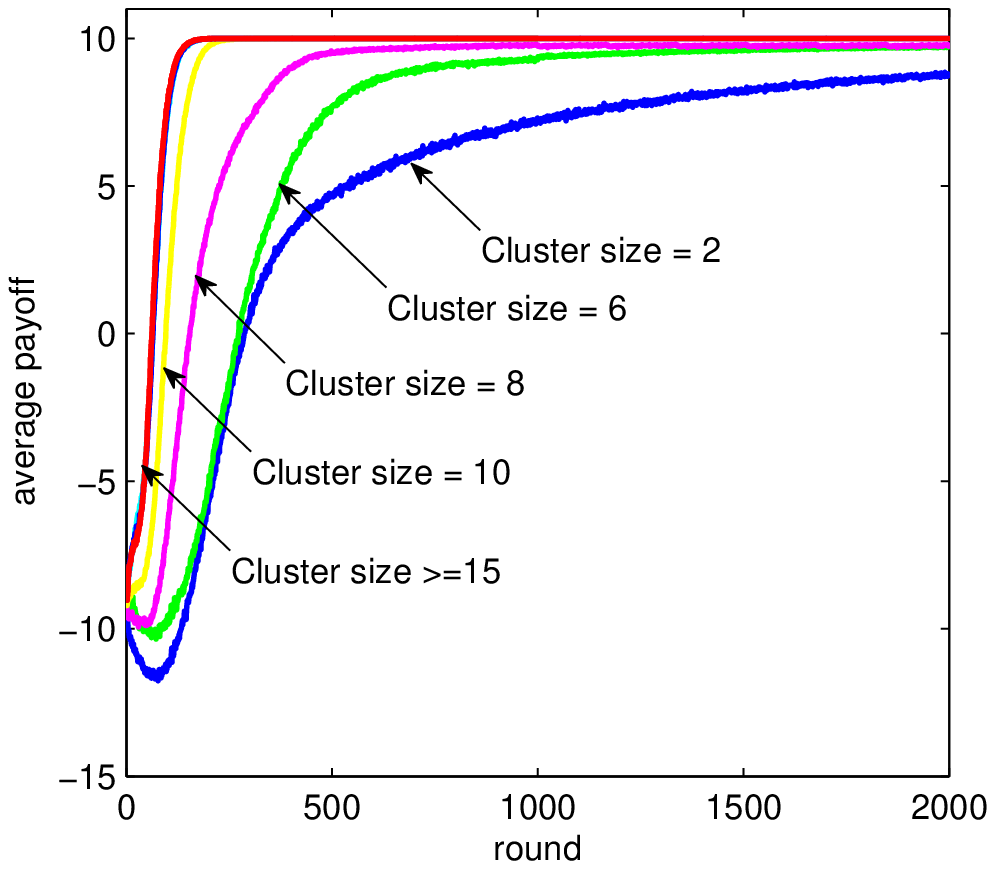}}
  	\caption{The influence of cluster size.} \label{figure9}
  \end{minipage}
  \hfill 
  \begin{minipage}[t]{.48\linewidth}
  	\centerline{\includegraphics[height = 2in, width = 2.4in]{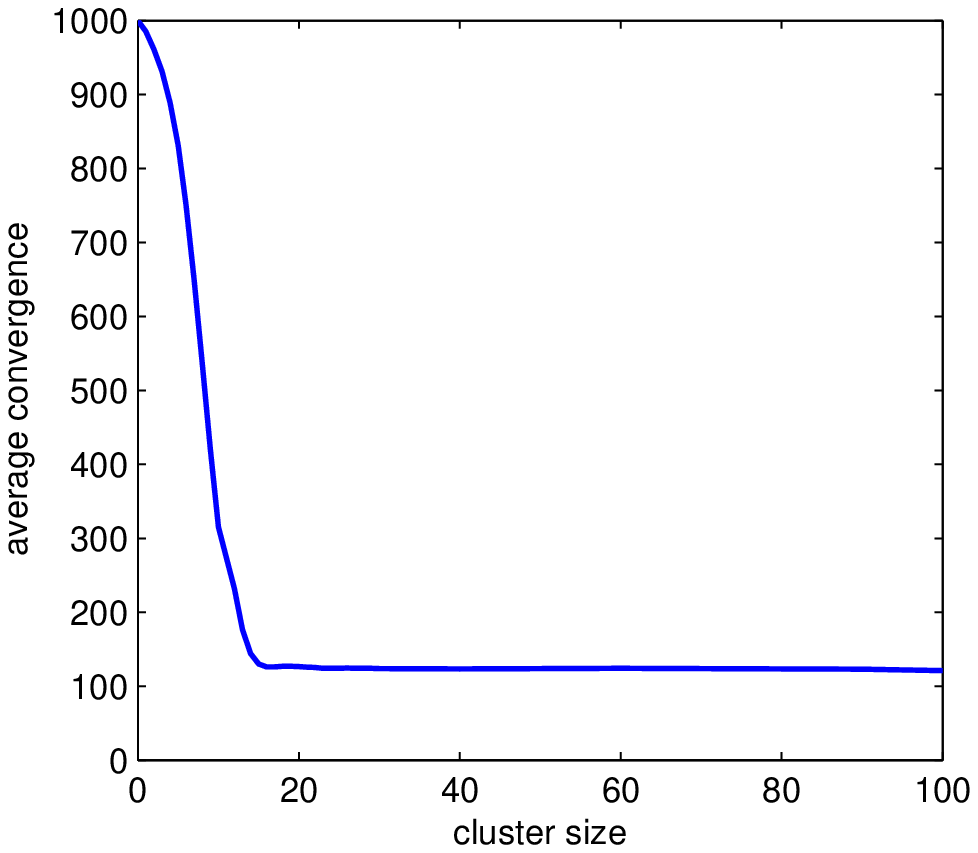}}
  	\caption{The average number of rounds needed before convergence under different cluster sizes.} \label{figure10}
  \end{minipage}
 \vfill
    \end{figure*}
\subsubsection{Influence of Cluster Size} \label{section4.5.1}
One unique feature of the hierarchical social learning framework is the division of clusters of agents. Fig \ref{figure9} and \ref{figure10} show the influence of cluster size on norm emergence with a population of 100 agents on the 6-action coordination game with high penalty (CGHP). From Fig \ref{figure9}, we can see that the norm emergence rate is gradually increased with the increase of cluster sizes, and stabilizes when the cluster size crosses 15. This phenomenon can be observed more clearly in Fig \ref{figure10}, which shows the average number of rounds needed before convergence is reduced with the increase of cluster size and stabilizes around 100 rounds. 

When the cluster size is increased to 100 in the extreme case, it is essentially reduced to centralized control in which only one supervisor agent supervises all the rest of agents. In this case, all the communication and computation burden would fall on this single supervisor agent. When the cluster size is 1, it is essentially equivalent with the case of the traditional social learning without a hierarchical structure. When the cluster size varies between 1 and 100, with the increase of the cluster size, each supervisor agent can supervise more subordinate agents and thus it is easier for agents to coordinate among each other. However, as the cluster size exceeds certain threshold, the advantage of centralized supervision diminishes. This property is desirable since the same level performance as fully centralized supervision can be achieved under distributed supervision, which not only increases the robustness of the HHLS and the framework itself but reduces the communication and computation burden of supervisor agents.

    \begin{figure*}
    \centering
  \begin{minipage}[t]{.5\linewidth}
  	\centerline{\includegraphics[height = 2in, width = 2.5in]{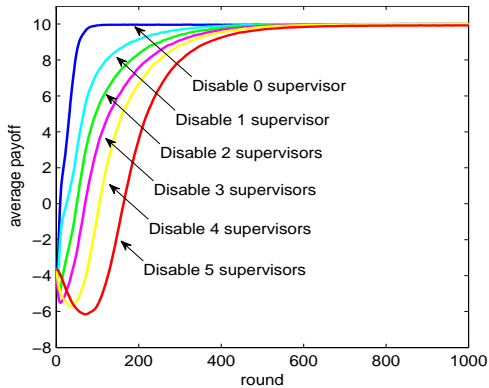}}
  	\caption{The influence of the number of disabled supervisors.} \label{figure11}
  \end{minipage}
  \vfill
    \end{figure*}
\subsubsection{Influence of The Number of Disabling Supervisors} \label{section4.5.2}
Next we further examine the robustness of HHLS in details by investigating the following questions: whether a consistent norm can still rapidly emerge and how is the emergence efficiency changed when a number of supervisors is disabled? Fig \ref{figure11} shows the dynamics of expected payoffs of agents with some supervisors disabled, and the results are averaged over 6-action coordination game with high penalty. We can see that hierarchically heuristic learning still enables agents to converge to a consistent norm when some of the supervisors is disabled. Though the convergence rate is gradually decreased with increased in the number of disabled supervisors, better performance than the traditional social learning framework can still be achieved. This is expected since those subordinate agents without supervisors can only learn based on their local information, and the hierarchical social learning framework would be reduced to the traditional social learning framework when all supervisors are disabled. 
\begin{figure*}
  \centering
  \begin{minipage}[t]{.46\linewidth}    
	\centerline{\includegraphics[height = 2in, width = 2.4in]{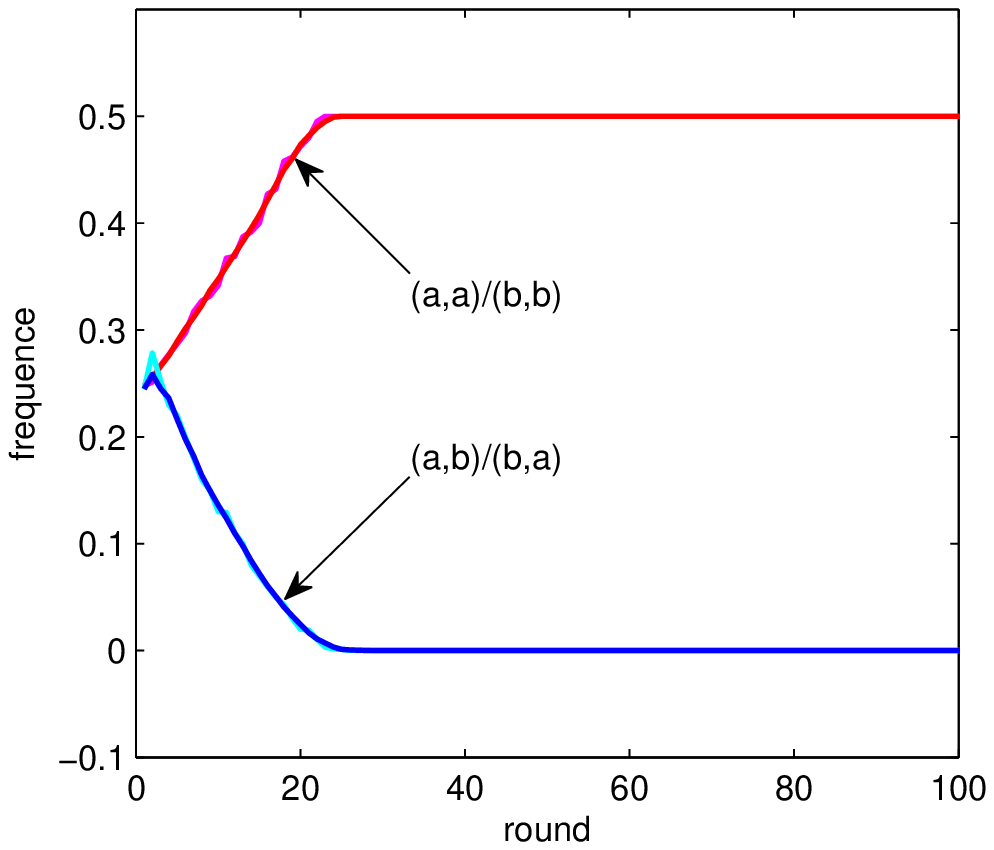}}
	\caption{The frequency of each joint action.} \label{figure12}
  \end{minipage}
  \hfill
  \begin{minipage}[t]{.47\linewidth}   
  	\centerline{\includegraphics[height = 2in, width = 2.4in]{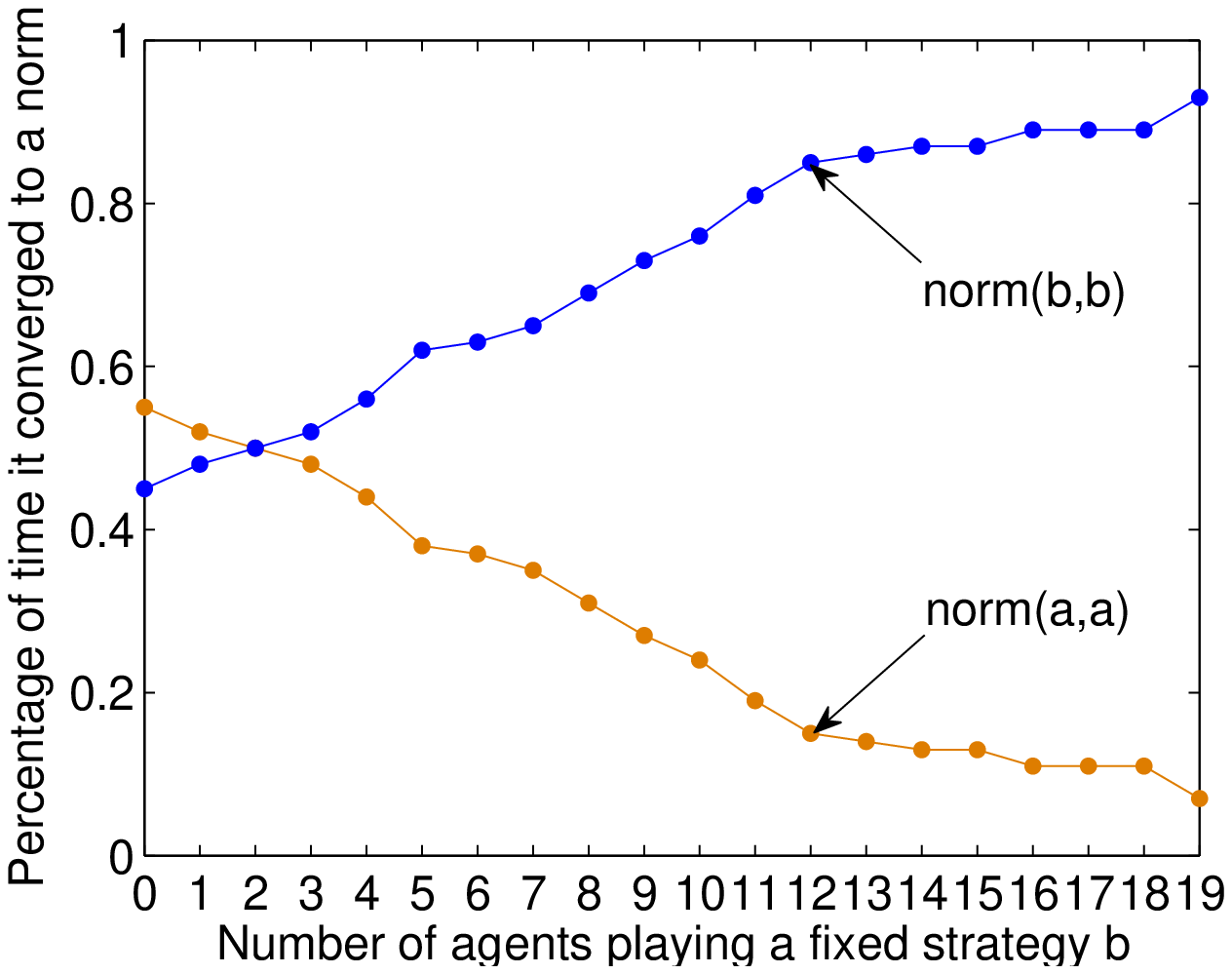}}
  	\caption{The influence of fixed-strategy agents on norm emergence.} \label{figure13}
  \end{minipage}
  \vfill
  \end{figure*}
\subsection{Fixed-strategy Agents on Norm Emergence} \label{section4.6}
In this section, we investigate the extraneous effects of norm emergence. Specifically, we consider injecting a small number of agents with a fixed-strategy, and study the influence of those fixed-strategy agents on the population's norm emergence. We compare the influence of fixed-strategy agents on HHLS with that on other learning strategies in previous work. Besides, we investigate the influence of different placement strategies of fixed-strategy agents on the convergence rate. For random networks, the location of each fixed-strategy agent is randomly selected. For other network topologies, the location of placing fixed-strategy agents could be critical in influencing the norm emergence in agent societies. We also consider the influence of late intervention and the influence of placing fixed-strategy agents on different levels within our hierarchical social learning framework.

\subsubsection{Influence of Fixed-strategy Agents on Norm Adoption} \label{section4.6.1}
When all agents are employing HHLS under the hierarchical social learning framework, for a 2-action coordination game, it can be observed that both norms ((a, a) and (b, b)) can be achieved with the same frequency over multiple runs as shown in Fig \ref{figure12}. This is understandable because the game itself is symmetric and the agents do not have any preference towards either of the two norms. 

In contrast, Fig \ref{figure13} shows the frequency of the two norms being evolved when the number of agents playing the fixed-strategy $b$ is increased gradually. Initially both norms ((a, a) and (b, b)) are evolved with roughly equal frequency. As the number of fixed-strategy agents increases, the frequency of evolving norm (b,b) gradually increases towards 1. This indicates that small amount of agents, using a fixed and deterministic strategy, can have significant influence on the overall population's norm emergence direction.

\subsubsection{Comparison with Previous Works}\label{section4.6.2}
\begin{figure*}
\centering
     \begin{minipage}[t]{.48\linewidth}
  	\centerline{\includegraphics[height = 2in, width = 2.4in]{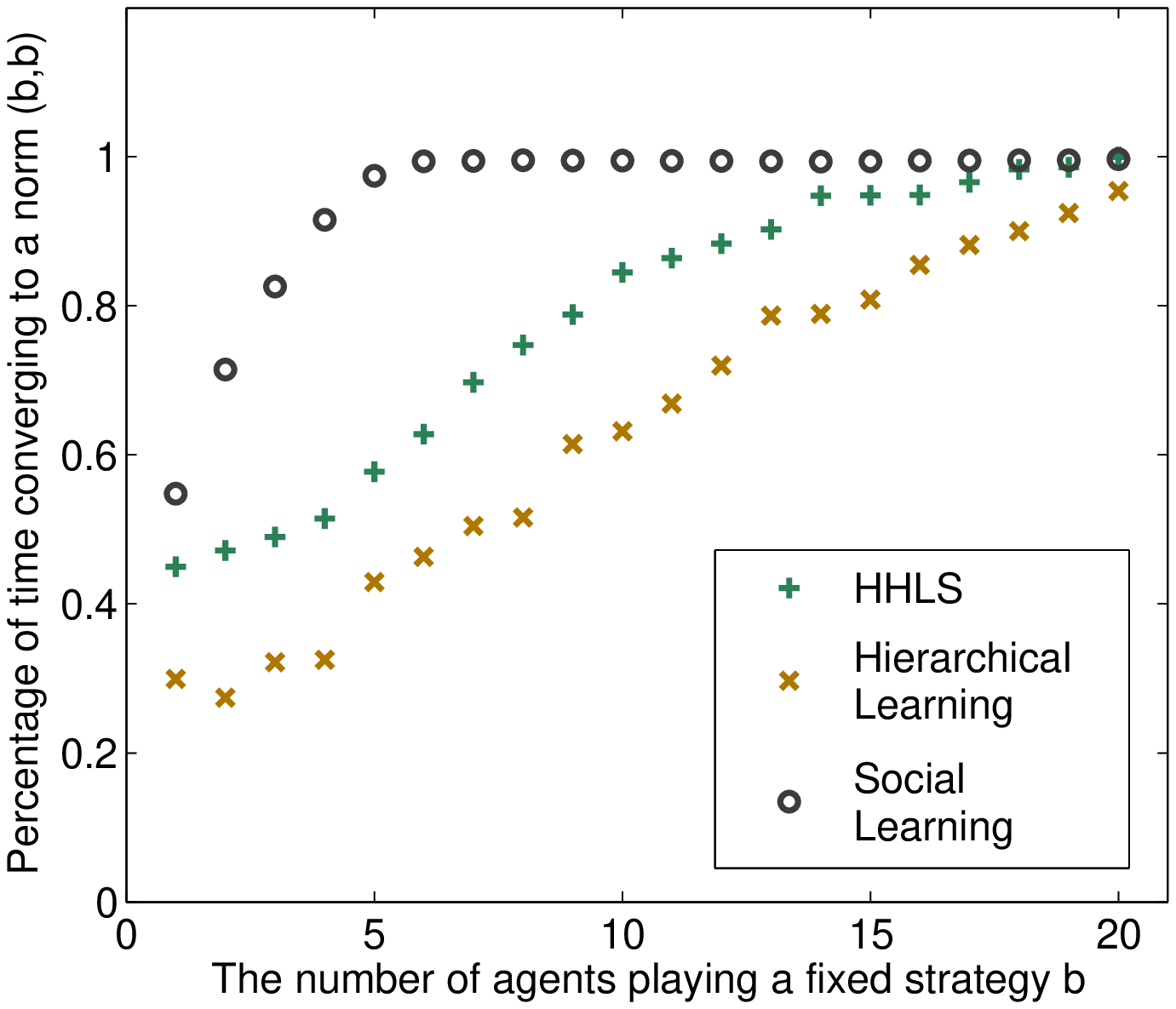}}
  	\caption{The influence of fixed-strategy agents on norm adoption under different strategies.} \label{figure17}
  \end{minipage}
  \hfill
   \begin{minipage}[t]{.48\linewidth}
  	\centerline{\includegraphics[height = 2in, width = 2.4in]{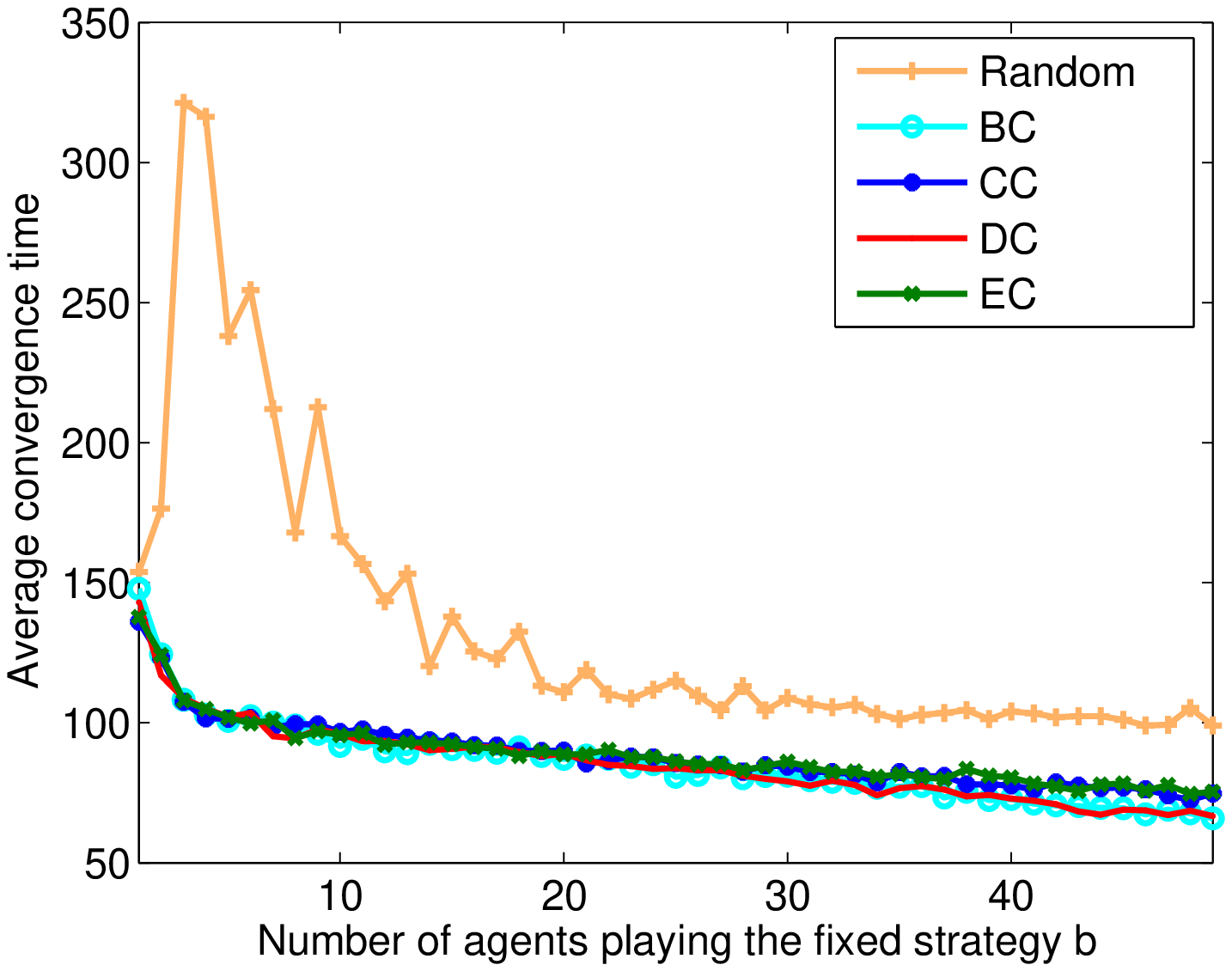}}
  	\caption{The influence of fixed-strategy agents on convergence time.} \label{figure15}
  \end{minipage}
  \vfill
\end{figure*}

Regarding the influence of fixed-strategy agent on norm adoption, we also compare our HHLS with social learning \cite{airiau2014emergence} and hierarchical learning \cite{yu2015hierarchical}. For HHLS and hierarchical learning \cite{yu2015hierarchical}, all fixed-strategy agents are inserted into groups of subordinate agents randomly. Fig \ref{figure17} shows the dynamics of the frequency of norm (b,b) emerging in 2-action coordination game as the number of fixed-strategy agents increases. The coordination game is symmetric: both norms ((a, a) and (b, b)) can be achieved with the same frequency. However, we see that hierarchical learning \cite{yu2015hierarchical} shows a slight bias towards norm (a,a) initially. Given the same number of fixed-strategy agents, agents are more inclined to learn the norm played by fixed-strategy agents under the social learning framework \cite{airiau2014emergence} than HHLS and hierarchical learning \cite{yu2015hierarchical}. This is because a hierarchical structure weakens the influence between subordinates and their neighbors through the supervisor's guidance.
\subsubsection{Influence of the Placement Strategies of Fixed-strategy Agents} \label{section4.6.3}
Previous sections investigate the effects of fixed-strategy agents which are randomly distributed in the population. Next we consider the influence of fixed-strategy agent placement strategies on the norm emergence rate. We are interested in investigating whether placing fixed-strategy agents following appropriate heuristics can decrease the convergence time. 

We consider the following commonly used placement metrics: degree (DC), betweenness (BC), closeness (CC) and eigenvector (EC) centrality heuristics. These metrics quantify the structural properties of the underlying network and previous work has shown these placement heuristics have better efficacy than random placement \cite{marchant2015manipulating,franks2013manipulating,griffiths2012impact}.  The DC of an agent is defined as the numbers of its neighbors. An agent with a higher DC means it can influence more agents directly in a network. The BC of an agent measures the number of the shortest paths in the network between other agents that pass through it. An agent with higher BC can manipulate the information flow more effectively in a network. The core idea of CC is that an agent is closeness central if it can interact all other agents easily. The closeness centrality of an agent located at a node is calculated using the average shortest path between the node and all other nodes. The EC of an agent is related to the connections of its neighbors. An agent has a higher EC if its neighbors has more connections with other agents. This measure takes into account both direct and indirect connections between agents. EC is calculated using the eigenvector of the largest eigenvalue given by the adjacency matrix representing the network. 

Fig \ref{figure15} shows the influence of placing fixed-strategy agents following the above placement metrics on convergence rate. Random placement is also considered as the baseline strategy. First we can see that the average convergence time is gradually decreased with the increase of the number of fixed strategy agents. Besides, BC, CC, DC and EC placement strategies outperform random placement. Intuitively fixed strategy agents located in nodes with higher centrality values can influence the whole population more quickly and widely, thus significantly reducing the convergence time compared with random placement. The performance difference between BC, CC, DC and EC is insignificant for scale-free network. This is due to the reason that almost the same set of agents are selected by these metrics in scale-free network since the values of Pearson's Correlation between these metrics are very high (approximately 0.9). 

\subsubsection{Late Intervention} \label{section4.6.4}
 
We have analyzed the influence of fixed-strategy agents on norm emergence where we place fixed-strategy agents at the beginning of the population interaction. Now we consider an alternative form of intervention, late intervention, i.e., inserting some number of fixed-strategy agents when a norm already emerges among the population, to investigate the robustness of our algorithm. We first allow all agents to employ HHLS without fixed-strategy agents being inserted. For a 3-action coordination game with high penalty (CGHP) presented in Table \ref{tableLateIntervation}, when the whole population has already converged to one of the optimal norms (e.g. norm (a,a)), we inject a small number of agents playing a fixed-strategy c. We expect these fixed-strategy agents to be able to influence the whole population to change to norm (c,c).

\begin{figure*}
\centering
    \begin{minipage}[t]{.48\linewidth}
  	\centerline{\includegraphics[height = 2in, width = 2.4in]{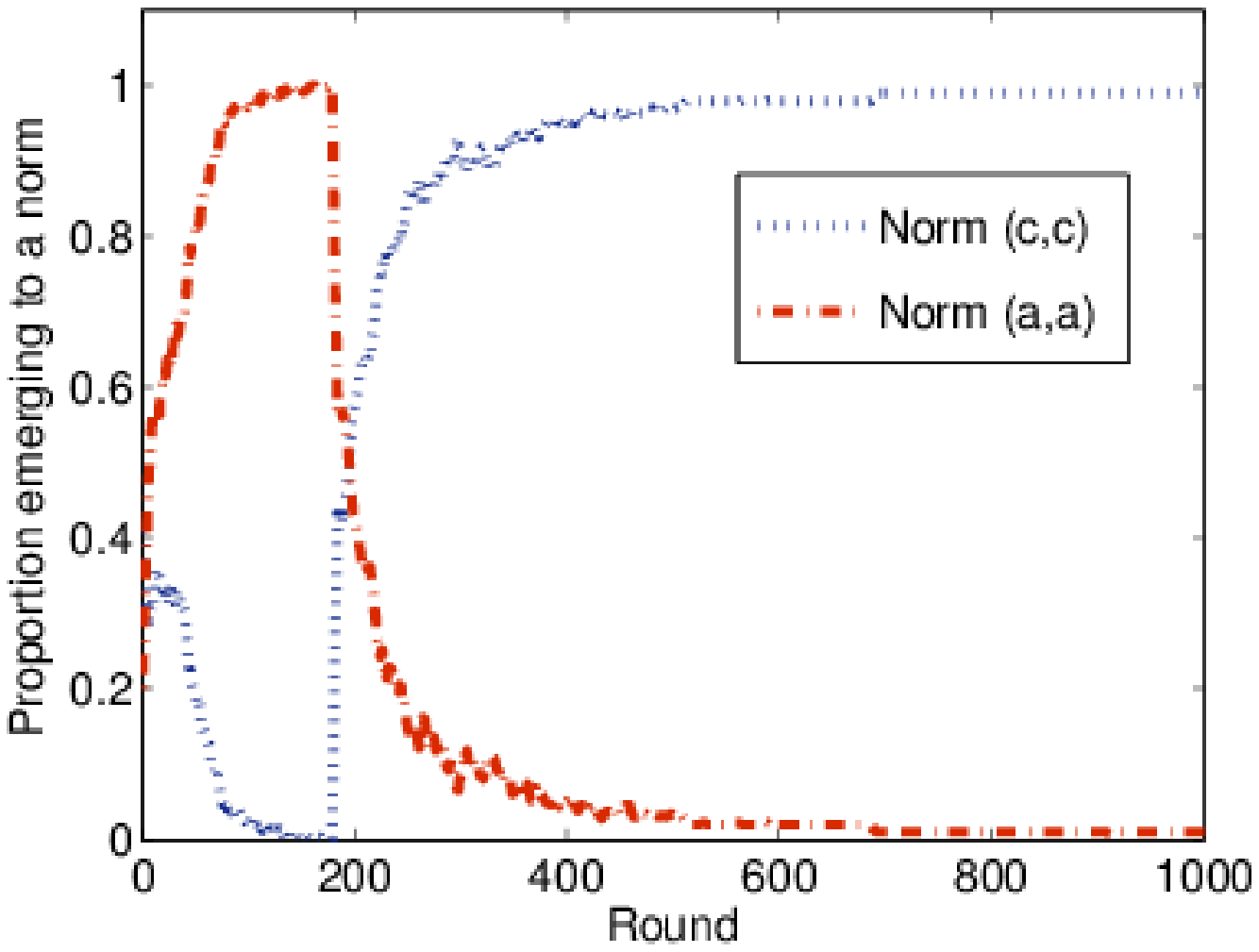}}
  	\caption{The influence of late intervention of fixed-strategy agents on norm emergence.} \label{figure23}
  \end{minipage}
  \hfill
  \begin{minipage}[t]{.46\linewidth}
  	\centerline{\includegraphics[height = 2in, width = 2.4in]{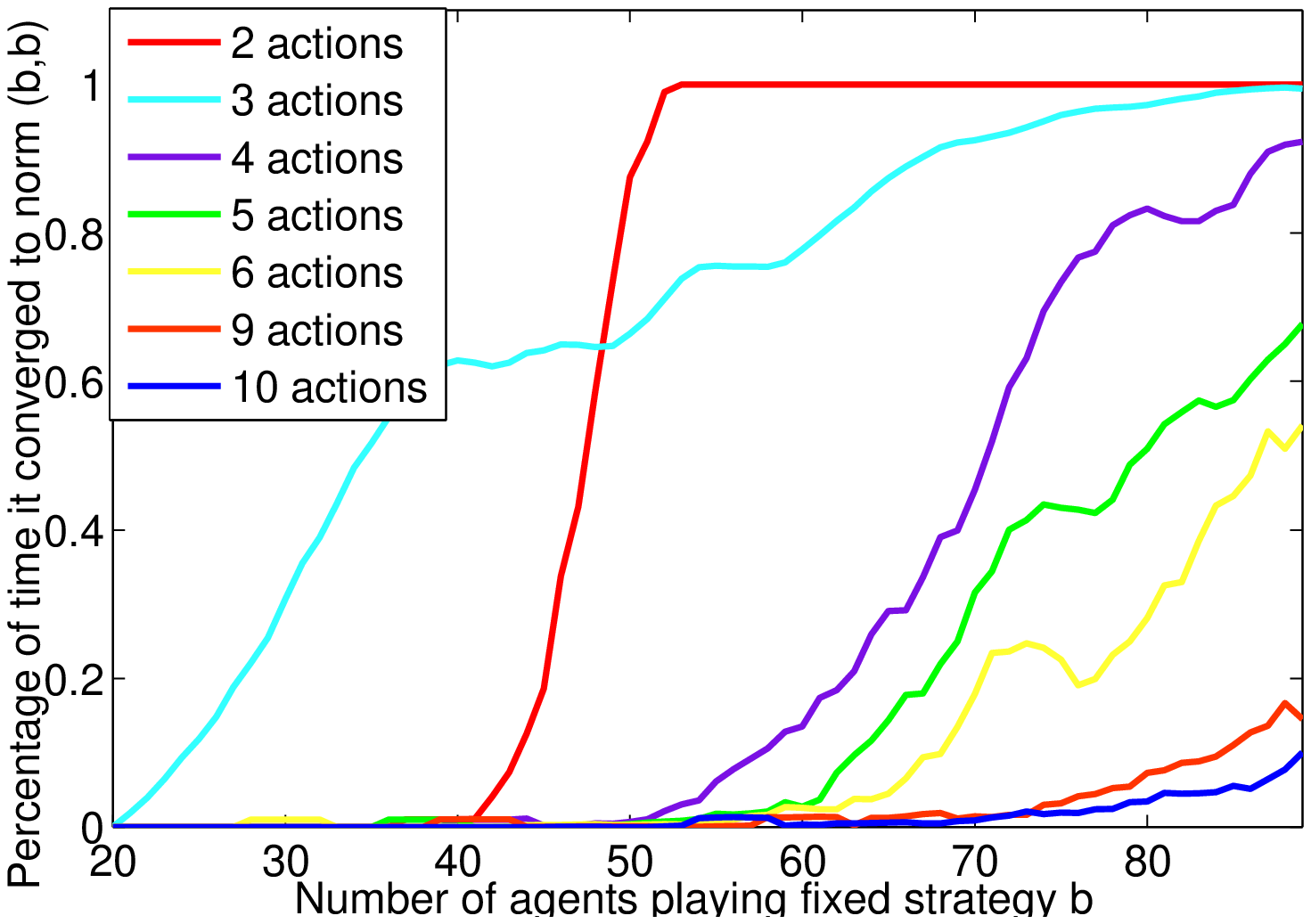}}
  	\caption{The influence of late intervention of fixed-strategy agents on norm emergence under different action sizes.} \label{figure22}
  \end{minipage}
    \vfill
\end{figure*}

Fig \ref{figure23} shows the influence of small number of fixed-strategy agents on replacing an existing norm with the 3-action coordination game with high penalty (CGHP). There exist two norms in CGHP and we can see that the whole population of 100 agents first converges to norm (a,a) without fixed-strategy agents at approximately the 190th time step. After this, we manually replace the top 20 agents with the highest EC values with fixed-strategy agents playing action c. We can see that the rest of agents are gradually incentivized to adopt strategy c and the whole population finally converges to norm (c,c) at approximately the 450th time step.

We further investigate the influence of late intervention using fixed-strategy agents on influencing an existing norm under different game sizes. The results are shown in Fig \ref{figure22}, which are averaged over 100 runs. We can see that for a 2-action coordination game, the frequency of norm (b,b) gradually increases towards 1 when the number of fixed-strategy agents is larger than 50. Compared with the results on initial intervention shown in Section \ref{section4.6.1}, we can see that late intervention needs a larger number of fixed-strategy agents to change the whole population's norm emergence direction. Another observation is that as the action space increases, late intervention of fixed-strategy agents fails to influence the whole population. This phenomenon indicates that the emerged norm becomes more stable with the increase of action space, in which late intervention may not exert any influence on norm emergence when the action space is sufficiently large. 

\begin{table}
\centering 
\caption{Coordination game with high penalty.} \label{tableLateIntervation}
\begin{tabular}{ccccc}
\hline 
\rule{0pt}{12pt}
\multirow{2}{0.5 in}{1's payoff \\ 2's payoff} &  & \multicolumn{3}{c}{Agent 2's actions} \\ 
\cline{3-5}
\rule{0pt}{12pt}
 & & a & b & c  \\ 
\hline
\multirow{3}{0.5 in}{\\Agent 1's \\ actions}\\
 & a & {\color{red}10} & 0 & -30 \\ 
\cline{2-5}
\rule{0pt}{12pt}
 & b & 0 & {\color{blue}7} & 0 \\ 
\cline{2-5}
\rule{0pt}{12pt}
 & c & -30 & 0 & {\color{red}10} \\ 
\hline 
	\end{tabular}
\end{table}

\subsubsection{Influence of Fixed-strategy Agents in Different Levels of the Hierarchical Framework} \label{section4.6.5}
\begin{figure*}
\centering
 \begin{minipage}[t]{.48\linewidth}
\centerline{\includegraphics[height = 2in, width = 2.5in]{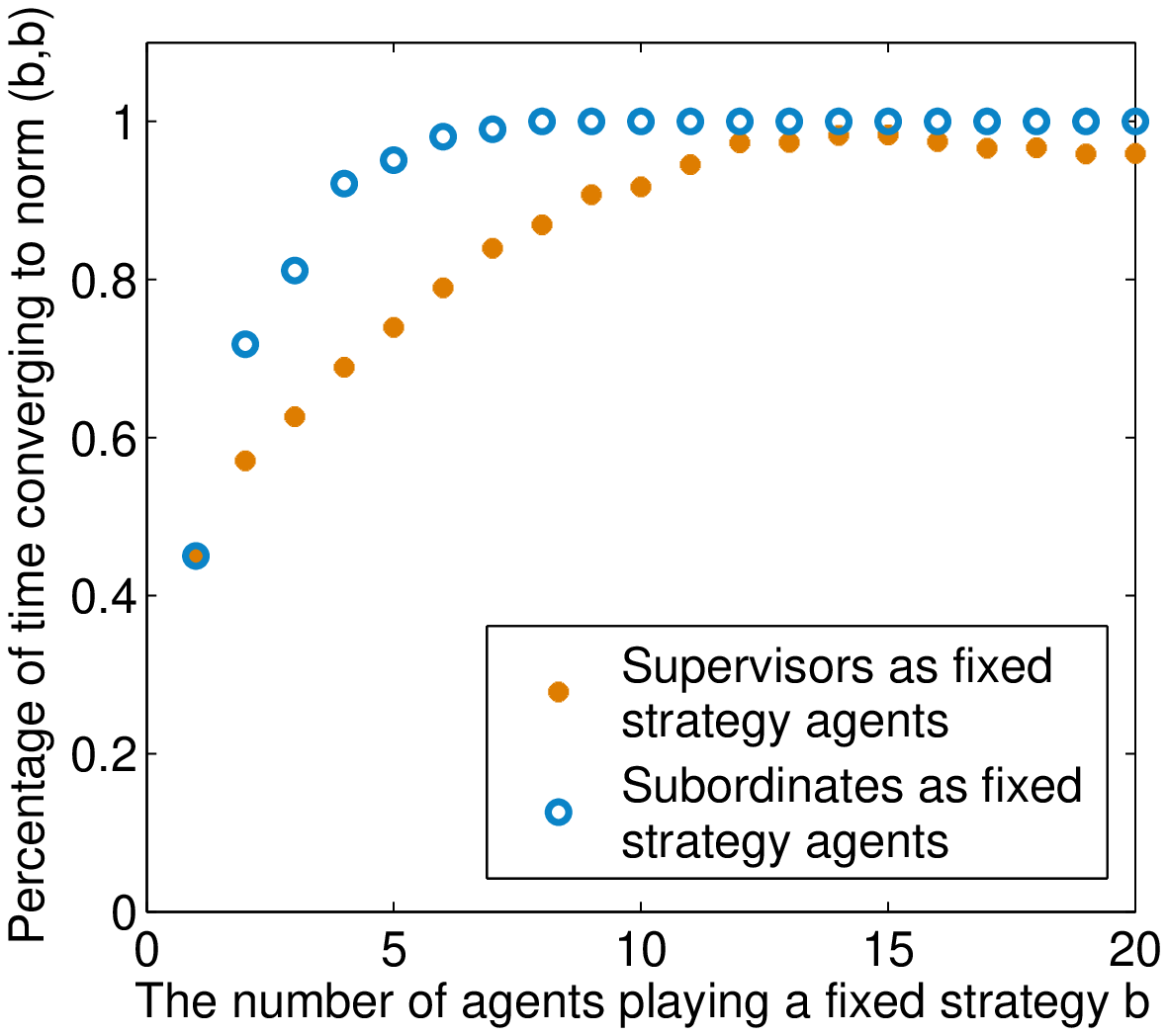}}
  	\caption{The influence of the location of fixed-strategy agents when they are placed in different levels.} \label{figure19}
  \end{minipage}
  \hfill
 \begin{minipage}[t]{.48\linewidth}
  	\centerline{\includegraphics[height = 2in, width = 2.5in]{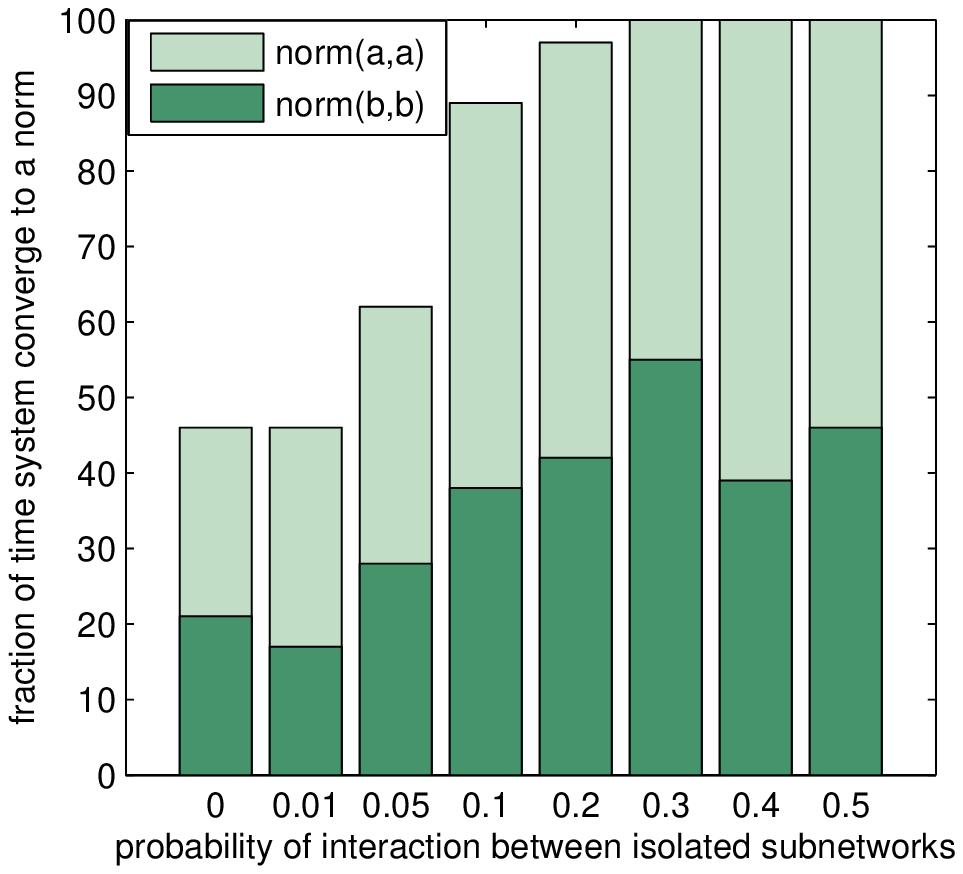}}
  	\caption{Norm emergence in isolated sub-networks.} \label{figure14}
\end{minipage}
\vfill
\end{figure*}

Finally, we investigate the influence of the location of fixed-strategy agents when they are placed in different levels (i.e. supervisor level and subordinate level) of the hierarchical learning framework. Fig \ref{figure19} shows the influence of the locations of fixed-strategy agents when they are placed in different levels given a 2-action coordination game. The whole population of 500 agents are divided into 20 groups using degree grouping. We can see that placing fixed-strategy agents in the supervisor level influences the whole population more quickly than placing fixed-strategy agents in the subordinate level. The result is reasonable because supervisors play a more important role than subordinates in the system. In addition to interacting with their neighbors as ordinary subordinate agents, supervisors collect all subordinates' information and generate recommendations for subordinates. With a number of supervisors playing a fixed-strategy, they can not only can influence their neighbors through interactions, but can pass their influence directly down to the group of subordinate agents they supervise through recommendations. Therefore, the whole system is more responsive towards the influence of fixed-strategy supervisor agents.

\subsection{Norm Emergence in Isolated Sub-networks} \label{section4.7}
In human societies, isolated populations can use contradictory norms, e.g., driving on the ``right" or the ``wrong" side of the road. When the frequency of interaction between the groups is quite small, a different norm can emerge in each group. To this end, we investigate the degree of isolation required for divergent norms to emerge in each group. 

We consider two groups of equal number of 100 agents and agents from one group interact with agents from the other group with certain probability. Results are presented in Fig \ref{figure14}. We observe that when the probability of interaction is higher than 0.2, a single norm pervades the entire population. In roughly half of the runs all agents learn to norm (a, a) and the other half of the runs, agents learn towards norm (b, b). However, when the probability of interaction becomes less than 0.2, there are runs where divergent norms emerge in two groups. It indicates that a higher interaction frequency between two isolated populations could sustain a consistent norm emergence for the whole system.

\section{Conclusion and Future Work} \label{section5}
We propose a hierarchically heuristic learning strategy to ensure efficient norm emergence in different distributed multiagent environments with various interaction topologies and several interaction scenarios (games). Extensive simulation shows that the proposed HHLS strategy can enable agents to reach consistent norms more efficiently and in a wider variety of games compared with previous approaches. We introduce different kinds of grouping mechanisms into our framework and evaluate the performance of these grouping mechanisms on norm emergence. The influence of different key non-hierarchical parameters (e.g., population size, action space, neighborhood size and network topology) are also investigated.  

We also evaluate the influence of centralized and decentralized hierarchically design by examining the effects of different cluster sizes (e.g., different number of supervisors). We find that certain degree of distributed supervision (a number of supervisors) can achieve the same performance as fully centralized supervision (only one supervisor), and thus make the HHLS robust towards the failure of certain supervisors. Moreover, we investigate the different components of the HHLS scheme and study the effect of the relative importance of those components. The results confirm that both components are critical and the integration of the two components leads to superior performance of HHLS. Furthermore, the influence of fixed-strategy agents on norm emergence are extensively investigated from different aspects and valuable insights are revealed (e.g., initial and late intervention, different placement strategies, fixed-strategy agents in different levels of the hierarchical structure). Lastly, we investigate the norm emergence in isolated population and experiments show that different norms emerge in isolated groups and a higher interaction frequency between isolated populations could sustain a consistent norm emergence for the whole system.


As future work, it is worthwhile investigating how to extend HHLS to multi-level hierarchical structures and how to design different communication and negotiation mechanisms between supervisors to accelerate norm emergence in larger action and agent spaces. Another worthwhile direction is to further investigate the influence of fixed-strategy agents in different ways, e.g., late intervention with some number of agents randomly selecting actions in contrast to fixed-strategy agents, mixing times between populations starting with different norms. 

\begin{acknowledgements}
This work is partially supported by the subproject of the National Key Technology R$\&$D Program of China (No.: 2015BAH52F01-1), National Natural Science Foundation of China (No.: 61304262, 61502072) and Tianjin Research Program of Application Foundation and Advanced Technology (No.: 16JCQNJC00100).

\end{acknowledgements}

\bibliographystyle{spmpsci}      
\bibliography{template}   

%
%

\end{document}